\documentstyle[epsfig,amssymb]{mn2e}
\def\simgt{\mathrel{\lower0.6ex\hbox{$\buildrel {\textstyle >}
 \over {\scriptstyle \sim}$}}}
\newcommand{\gtsim}{\mbox{{\raisebox{-0.4ex}{$\stackrel{>}{{\scriptstyle\sim}}
$}}}}
\newcommand{\ltsim}{\mbox{{\raisebox{-0.4ex}{$\stackrel{<}{{\scriptstyle\sim}}
$}}}}
\newcommand{\mc}{\multicolumn}
\def \kband{$K$-band }
\def \rband{$R$-band }

\begin{document}

\title[The 6C** sample II: Redshift distribution]{The 6C** sample of
  steep-spectrum radio sources: II -- Redshift distribution and the
  space density of high-redshift radio galaxies}

\author[Maria J.~Cruz et al.]
{Maria J.~Cruz$^{1,2}$\thanks{Email: mjc@astro.ox.ac.uk}, Matt J.~Jarvis$^{1}$, Steve Rawlings$^{1}$ and Katherine M.~Blundell$^{1}$\\
\footnotesize
\\
$^{1}$Astrophysics, Department of Physics, Keble Road, Oxford, OX1 3RH, UK \\
$^{2}$Leiden University, Sterrewacht, Oort Gebouw, P.O. Box 9513, 2300 RA Leiden, The Netherlands
}

\maketitle

\begin{abstract}
We use the 6C** sample to investigate the co-moving space density of
powerful, steep-spectrum radio sources. This sample, consisting of 68
objects, has virtually complete \kband photometry and spectroscopic
redshifts for 32 per cent of the sources. In order to find its
complete redshift distribution, we develop a method of redshift
estimation based on the $K-z$ diagram of the 3CRR, 6CE, 6C* and 7CRS
radio galaxies.  Based on this method, we derive redshift probability
density functions for all the optically identified sources in the 6C**
sample. Using a combination of spectroscopic and estimated redshifts,
we select the most radio luminous sources in the sample. Their
redshift distribution is then compared with the predictions of the
radio luminosity function of Jarvis et al.  We find that, within the
uncertainties associated with the estimation method, the data are
consistent with a constant co-moving space density of steep-spectrum
radio sources beyond $z \,\,\gtsim\,\, 2.5$, and rule out a steep
decline.

\end{abstract}

\section{INTRODUCTION}

Powerful radio sources, such as radio galaxies and quasars, trace the
most massive galaxies (Jarvis et al. 2001a; De Breuck et al. 2002;
Willott et al. 2003; Zirm, Dickinson \& Dey 2003) and are associated
with the most massive black holes (Dunlop et al. 2003; McLure et
al. 2004; McLure \& Jarvis 2004) in the Universe, at all cosmic
epochs.  Radio galaxies have been detected up to redshifts of just
above five (van Breugel et al. 1999), and quasars up to and beyond
redshifts of six (Fan et al. 2003), leaving little time during which
quasars and their host galaxies could form.  This provides a challenge
for hierarchical galaxy formation models, though some recent
semi-analytic models are able to produce significant numbers of
massive galaxies at high redshifts (e.g. Bower et al.\ 2006; Night et
al.\ 2006).  Constraining the space-density of high-redshift radio
sources is therefore important, as it has implications on the theories
of galaxy and structure formation. Using radio sources is also
advantageous in this respect because they are selected on the basis of
their radio emission and are thus free of the problems associated with
optical selection methods, such as dust obscuration (cf. optically
selected quasars).

 It is clear that the co-moving space densities of the rarest, most
 powerful radio sources were much higher around $z \sim 2$ than they
 are at present (Longair 1966), but the form of the evolution beyond
 that redshift is still a matter of debate. The question of a
 `redshift cut-off' (Dunlop \& Peacock 1990) in the radio source
 population has come under careful scrutiny in recent years (Jarvis \&
 Rawlings 2000; Jarvis et al. 2001c).  The current situation is one in
 which there is no compelling evidence of a high-redshift decline in
 low-frequency selected (i.e. predominantly steep-spectrum) radio
 sources, whereas there is evidence for a slight decline in radio-loud
 quasars from high-frequency selected (i.e. predominantly
 flat-spectrum) samples (Dunlop \& Peacock 1990, Shaver et al. 1996;
 Jarvis \& Rawlings 2000; Wall et al. 2005).

In this study we focus on the low-frequency selected population
(predominantly radio galaxies). To this effect, we use a new sample of
radio sources drawn from the 151~MHz 6C survey, which has been
filtered with radio criteria designed to optimize the chances of
finding radio galaxies at $z > 4$.  This sample, namely 6C**, has been
selected to be brighter than 0.5~Jy at 151~MHz. Additional selection
criteria have excluded all sources with radio spectral index between
151~MHz and 1.4~GHz flatter than 1, or with radio angular size larger
than 13~arcsec. The final sample consists of 68 objects over an area
of sky of 0.421~sr, and is statistically complete at an angular size
limit of $\theta < 11$~arcsec. Full details of how the 6C** sample was
selected can be found in Cruz et al. (2006, hereafter Paper I).

The selection criteria just described are similar to those of the 6C*
sample (Blundell et al. 1998; Jarvis et al. 2001a,b), which was one of
the samples used by Jarvis et al. (2001c) to constrain the co-moving
space density of low-frequency selected radio sources. The 6C* sample
was crucial in that study in sampling to high redshift ($z \simeq
4.4$).  The 6C** sample, being larger (cf. 0.13~sr) and deeper
(cf. $0.96 \leq S_{151} \leq 2.0$~Jy) than 6C*, aims to improve on the
small-number statistics limitation of this previous work, and
ultimately to extend it to higher redshifts ($z \,\,\gtsim\,\, 5$).

Deep \kband imaging follow-up with UFTI/UIST on UKIRT, NIRI on Gemini
and NIRC on Keck provided photometry for all members of the 6C**
sample (Paper I). Optical spectroscopy provided redshifts for 32 per
cent of the sources (Paper I and references therein). A summary of key
observational information is given in
Table~\ref{tab:6cssummary1_median}.

\begin{table*}
\scriptsize
\begin{center}
\begin{tabular}{lcccllcl}
\hline
\mc{1}{c}{(1)} & \mc{1}{c}{(2)} & \mc{1}{c}{(3)} & \mc{1}{c}{(4)} &
\mc{1}{c}{(5)} & \mc{1}{c}{(6)} & \mc{1}{c}{(7)} & \mc{1}{c}{(8)} \\
\mc{1}{c}{Source} & \mc{1}{c}{$S_{151}$} &
\mc{1}{c}{$\alpha^{1400}_{151}$} & \mc{1}{c}{$K$} & \mc{1}{c}{$z$} &
\mc{1}{c}{Line} & 
\mc{1}{c}{$\log_{10}L_{\rm line}$} & \mc{1}{c}{Ref.}\\
\hline
6C**0714+4616 & 1.65 & 1.25  & 16.316(8) &  1.466  & CIV & 36.05 & Cea\\
6C**0717+5121 & 1.24 & 1.11  & 17.788(5) &  nd     &\\
6C**0726+4938 & 0.61 & 1.19  & 18.168(8) &  1.203? & [O II]? & 35.15&
Cea\\
6C**0737+5618 & 0.74 & 1.28  & $>21.1$(8)&  nd     &\\        
6C**0744+3702 & 0.64 & 1.31  & 19.440(8) &  2.992  & Ly$\alpha$  &
35.78 & DBea\\
\\
6C**0746+5445 & 0.53 & 1.04  & 18.423(3) & 2.156 & Ly$\alpha$ & -- & Cea\\
6C**0754+4640 & 0.69 & 1.07  & 19.971(5) & nd    &\\
6C**0754+5019 & 1.05 & 1.07  & 20.629(5) & 2.996 & Ly$\alpha$ &
35.81 & Cea\\
6C**0801+4903 & 1.08 & 1.13  & 19.855(5) &       &\\
6C**0810+4605 & 10.26& 1.01  & 15.993(8) & 0.620 & [O II] & 36.17 &
Cea\\
\\
6C**0813+3725 & 0.50 & 1.24  & 18.798(8) & nd    & \\
6C**0824+5344 & 0.88 & 1.06  & 19.392(8) & 2.824 & Ly$\alpha$ & 36.94
& Cea\\
6C**0829+3902 & 0.51 & 1.16  & 19.413(5) & nd    & \\
6C**0832+4420 & 0.52 & 1.14  & 18.915(8) &       & \\
6C**0832+5443 & 0.60 & 1.02  & 19.283(5) & 3.341 & Ly$\alpha$ & 36.68
& Cea\\
\\
6C**0834+4129 & 0.50 & 1.00  & 19.378(8) & 2.442 & Ly$\alpha$ & 36.35
& Cea\\
6C**0848+4803 & 0.71 & 1.26  & 17.828(8) &       &\\
6C**0848+4927 & 0.94 & 1.03  & 18.222(8) & nd    &\\
6C**0849+4658 & 3.50 & 1.03  & 17.319(8) &       &\\
6C**0854+3500 & 0.87 & 1.06  & 18.121(8) & 2.382 & Ly$\alpha$ & -- &
Cea\\
\\
6C**0855+4428 & 0.94 & 1.06  & 18.485(5) &       &\\
6C**0856+4313 & 0.59 & 1.00  & 17.999(8) & 1.761 & Ly$\alpha$ & 36.44
& Cea\\
6C**0902+3827 & 1.60 & 1.04  & 19.290(5) & nd    &\\
6C**0903+4251 & 3.14 & 1.08  & 16.615(8) & 0.907 & [O II] & 35.20 & McC\\
6C**0909+4317 & 3.36 & 1.01  & 18.635(8) &       &\\
\\
6C**0912+3913 & 0.56 & 1.02  & 17.595(8) &       &\\
6C**0920+5308 & 0.56 & 1.03  & 14.526(8) &       &\\
6C**0922+4216 & 2.70 & 1.06  & 15.928(8) & 1.750 & & & Vea\\
6C**0924+4933 & 0.93 & 1.05  & 14.955(8) &       &\\
6C**0925+4155 & 0.91 & 1.01  & 20.279(5) & nd    &\\
\\
6C**0928+4203 & 2.04 & 1.22  & 18.448(8) & 1.664 & Ly$\alpha$ & 36.95
& Cea\\
6C**0928+5557 & 0.58 & 1.04  & 17.285(5) &       &\\
6C**0930+4856 & 0.66 & 1.02  & 18.903(8) &       &\\
6C**0935+4348 & 1.09 & 1.36  & $>20.9$(8)& 2.321? & Ly$\alpha$? & 36.70
& Cea\\
6C**0935+5548 & 0.90 & 1.01  & 18.325(8) &       &\\
\\
6C**0938+3801 & 1.03 & 1.13  & 18.132(8) & nd    &\\
6C**0943+4034 & 0.99 & 1.06  & 17.592(8) &       &\\
6C**0944+3946 & 0.66 & 1.00  & 19.088(8) &       &\\
6C**0956+4735 & 6.13 & 1.13  & 17.192(8) & 1.026 & [O II] &
35.75 & McC\\
6C**0957+3955 & 0.62 & 1.01  & 18.264(8) &       &\\
\\
6C**1003+4827 & 6.88 & 1.09 & 16.950(8) &       &\\
6C**1004+4531 & 0.70 & 1.01 & 17.183(8) &       &\\
6C**1006+4135 & 0.52 & 1.01 & 19.825(5) &       &\\
6C**1009+4327 & 2.89 & 1.23 & 20.513(3) & 1.956 & Ly$\alpha$ & 35.64
& Cea\\
6C**1015+5334 & 1.44 & 1.05 & 18.516(8) &       &\\
\\
6C**1017+3436 & 1.17 & 1.04 & 18.972(8) &       &\\
6C**1018+4000 & 0.53 & 1.02 & 18.434(8) &       &\\
6C**1035+4245 & 1.89 & 1.28 & 17.250(8) &       &\\
6C**1036+4721 & 3.70 & 1.03 & 16.967(8) & 1.758 & Ly$\alpha$ &
-- & Cea\\ 
6C**1043+3714 & 2.62 & 1.04 & 17.579(3) & 0.789 &  & & ASea\\
\\
6C**1044+4938 & 1.66 & 1.08 & 18.685(5) &       &\\
6C**1045+4459 & 0.95 & 1.11 & 18.438(5) & 2.571 & Ly$\alpha$ & 36.55
& Cea\\
6C**1048+4434 & 1.51 & 1.02 & 18.628(5) &       &\\
6C**1050+5440 & 0.93 & 1.20 & 19.715(8) &  nd   &\\
6C**1052+4349 & 0.51 & 1.03 & 17.081(8) &       &\\
\\
6C**1056+5730 & 2.66 & 1.12 & 17.295(8) &       &\\
6C**1100+4417 & 0.72 & 1.09 & 18.095(8) &       &\\
6C**1102+4329 & 1.11 & 1.08 & 19.661(8) & 2.734 & Ly$\alpha$ & 36.62
& Cea\\
6C**1103+5352 & 2.67 & 1.03 & 20.142(5) & \\
6C**1105+4454 & 0.83 & 1.01 & 17.729(8) & \\
\\
6C**1106+5301 & 0.77 & 1.13 & 17.354(8) & \\
6C**1112+4133 & 0.54 & 1.33 & 18.044(8) & \\
6C**1125+5548 & 0.63 & 1.23 & 19.680(5) & \\
6C**1132+3209 & 0.63 & 1.04 & 14.505(3) & 0.231 & & &Bea\\
6C**1135+5122 & 0.66 & 1.10 & 18.554(5) &      \\
\hline
\end{tabular}
\end{center}
\end{table*}

\begin{table*}
\scriptsize
\begin{center}
\begin{tabular}{lcccllcl}
\hline
\mc{1}{c}{(1)} & \mc{1}{c}{(2)} & \mc{1}{c}{(3)} & \mc{1}{c}{(4)} &
\mc{1}{c}{(5)} & \mc{1}{c}{(6)} & \mc{1}{c}{(7)} & \mc{1}{c}{(8)} \\
\mc{1}{c}{Source} & \mc{1}{c}{$S_{151}$} &
\mc{1}{c}{$\alpha^{1400}_{151}$} & \mc{1}{c}{$K$} & \mc{1}{c}{$z$} &
\mc{1}{c}{Line} & 
\mc{1}{c}{$\log_{10}L_{\rm line}$} & \mc{1}{c}{Ref.}\\
\hline	     				 				
6C**1138+3309 & 0.93 & 1.22 & 18.014(8) &      \\
6C**1138+3803 & 0.51 & 1.05 & 17.351(3) &      \\
6C**1149+3509 & 0.61 & 1.06 & 18.729(8) &      \\
\hline
\end{tabular}
\end{center}
{\caption{\label{tab:6cssummary1_median} Summary of the observational
data on the 6C** sample.  {\bf Column 1:} Name of the 6C**
source. {\bf Column 2:} 151~MHz flux-density measurements in Jy from
the 6C survey (Hales et al. 1988; Hales et al. 1990).  {\bf Column 3:}
Radio spectral index evaluated between 151~MHz and 1.4~GHz from the 6C
and NRAO VLA SKy Survey (NVSS; Condon et al. 1998) flux densities.
{\bf Column 4:} \kband magnitude within the angular aperture in arcsec
given in brackets.  {\bf Column 5:} Redshift: `?' signifies that this
value is uncertain, 'nd' signifies that the source was observed but
there were no emission lines and/or continuum detected.  {\bf Column
6:} Prominent emission line in the existing spectra, `?' signifies
that the line identification is uncertain.  {\bf Column 7:}
$\log_{10}$ of the luminosity of the line listed in Column 6 (measured
in units of W): `--' signifies that the data were inadequate to obtain
a line luminosity, due to the absence of a spectrophotometric standard
or non-photometric observing conditions.  {\bf Column 8:} Reference
for the redshift of the source, Cea $=$ Cruz et al. (2006), DBea $=$
De Breuck et al. (2001), McC $=$ McCarthy (1991), Vea $=$ Vigotti et
al. (1990), ASea $=$ Allington-Smith et al. (1985), Bea $=$ Brinkmann
et al. (2000).}}
\end{table*}

In this paper we describe a method of redshift estimation based on the
$K-z$ diagram of radio galaxies. This is presented in
Section~\ref{sec:estimation}. In Section~\ref{sec:estimates}, we use
the complete set of \kband magnitudes of the 6C** sample to estimate
redshifts for all its optically identified members. These are compared
to spectroscopic redshifts in Section~\ref{sec:comparison}, in order
to assess the robustness of the method. The resulting estimated
redshift distribution is discussed in Section~\ref{sec:est-zdist}.  In
Section~\ref{sec:JarvisRLF} we summarize the model radio luminosity
function (RLF) of Jarvis et al. (2001c). This is the most relevant
model to compare our data to, because it takes into account the
selection effects of the 6C* sample.  In Section~\ref{sec:RLF} we
compare the redshift distribution (including spectroscopic redshifts)
of the 6C** sample with the model predictions, and discuss the
evolution of the co-moving space density of the most radio luminous,
low-frequency selected sources.  Unless otherwise stated, we assume
throughout that $H_{0}=70~ {\rm km~s^{-1}Mpc^{-1}}$, $\Omega_ {\mathrm
M} = 0.3$ and $\Omega_ {\Lambda} = 0.7$. The convention used for radio
spectral index is $S_{\nu} \propto \nu^{-\alpha}$, where $S_{\nu}$ is
the flux-density at frequency $\nu$.

\section{Redshift Estimation Method}\label{sec:estimation}

Infrared-photometry provides a method of redshift estimation by
utilising the tightness of the relation between \kband magnitude and
redshift, which is characteristic of the near-infrared Hubble diagram
of radio galaxies (Lilly \& Longair 1984; Eales et al. 1997; Jarvis et
al.  2001a; De Breuck et al. 2002, Willott et al. 2003).  The physical
basis for the $K-z$ relation is not well understood.  At low
redshifts, the \kband emission is dominated by the old stellar
population in the host galaxy; at high redshifts, \kband
samples rest-frame optical wavelengths, where the star formation
history can have a significant effect. Non-stellar contamination to
the \kband light, in the form of reddened quasar light and/or narrow
emission lines, also contributes to the difficulty of interpreting the
$K-z$ diagram of radio galaxies, particularly at high redshifts ($z >
3$). Despite these caveats, the $K-z$ diagram is still of interest as
a tool for redshift estimation.

Redshift estimates based on the $K-z$ diagram have generally been
obtained by simple application of the empirical $K-z$ relation
(e.g. Dunlop \& Peacock 1990).  However, the significant amount of
scatter around this relation requires the use of a more sophisticated
method -- one which takes into account all the available information
in the diagram, and also which allows us to characterise the
uncertainty on the output redshift estimates. With these requirements
in mind the following approach is adopted: (i) we use Monte Carlo
simulations to generate a statistical universe of synthetic
realisations of the $K-z $ diagram, based on a model of its underlying
galaxy distribution, and (ii) we extract individual photometric
redshift probability density functions from this simulated population.

\subsection{The $K-z$ Diagram for the 3CRR, 6CE, 6C* and 7CRS radio galaxies
}
\label{sec:kz_willott}

The most well defined $K-z$ diagram for radio galaxies currently
available is the one obtained by Willott et al. (2003) from a combined
dataset of the radio galaxies from the 3CRR (Laing, Riley \& Longair
1983), 6CE (Eales et al. 1997; Rawlings, Eales \& Lacy 2001), 6C*
(Jarvis et al. 2001a,b) and 7CRS (Lacy et al. 2000, Willott et
al. 2003) flux-limited samples.  It is based on a total of 204 radio
galaxies with redshifts ranging from 0.05 to 4.4, and its $K-z$ relation
is well fitted by a second-order polynomial between $K$-magnitude and
$\log_{10} z$ (Willott et al. 2003):
\begin{equation}
K(z) = 17.37 + 4.53\,\log_{10}\, z - 0.31\,(\log_{10}\, z)^2. 
\label{eq:K-z}
\end{equation}
The main advantage of using this $K-z$ diagram is that it has been
obtained from completely identified samples with close to complete, or
complete redshift information. This ensures the absence of significant
biases in terms of sources with the weakest lines being missed because
their redshifts are difficult to obtain.  Another advantage is that
these samples have been selected at a similar radio-frequency to 6C**,
with progressively fainter flux-density limits.  The brightest sample
is 3CRR selected at 178\,MHz, with a flux-density limit of $S_{178}
\geq 10.9$\,Jy ($S_{151} \geq 12.4$\,Jy, assuming a spectral index of
0.8); the faintest sample is 7CRS selected at 151\,MHz, with a
flux-density limit of $S_{151} \geq 0.5$\,Jy. The intermediate samples
are 6CE and 6C* selected at 151\,MHz, with flux-density limits of $2.0
\leq S_{151} \leq 3.93$\,Jy and $0.96 \leq S_{151} \leq 2.00$\,Jy,
respectively.  This results in a wide range in radio luminosity, which
has made the investigation of the radio-luminosity dependence of the
$K-z$ relation possible in an unprecedented way.  The correlation
between \kband luminosity and radio luminosity has been one of the
major worries with redshift estimates based on the $K-z$ relation.

Willott et al. (2003) found a statistically significant mean
luminosity difference between the 3CRR and 7CRS radio galaxies of
0.55~mag in $K$-band, over all redshifts. However, the 6C radio
galaxies were found to differ on average from the 3C ones by only
$\simeq 0.3$~mag, which is much smaller than the value ($\simeq
0.6$~mag) reported previously (Eales et al. 1997).  The mean
luminosity difference between the 6C and 7C galaxies was not found to
be significant.  These results are confirmed by McLure et al. (2004),
who used HST data to study the host galaxy properties of a sample of
radio galaxies at $0.4 \leq z \leq 0.6$, which spans three decades in
radio luminosity. They found mean luminosity differences in the \rband
of $\simeq 0.3$~mag between the 3C and 6C samples, and $\simeq 0.8$~mag
between the 3C and 7C samples\footnote{These magnitude differences
were measured in the $I$-band and converted to the \rband assuming a
constant $R - I$ colour (McLure et al. 2004).}. These results are in
good quantitative agreement with those of Willott et al. (2003).  From
both studies it follows that there is a correlation between radio
luminosity and host luminosity within the 3CRR, 6CE and 7CRS
samples. This has been interpreted as suggestive of a correlation of
both these properties with black hole mass (Willott et al. 2003;
McLure et al. 2004).  However, the weakness of these correlations
means that the radio galaxies within the combined dataset used by
Willott et al. (2003) follow essentially the same $K-z$ relation.

\subsection{Parametric modelling of the $K-z$ Diagram}
\label{sec:kz_model}

\begin{figure}
\begin{center} 
{\hbox to 0.45\textwidth{ \epsfxsize=0.5\textwidth \epsfbox{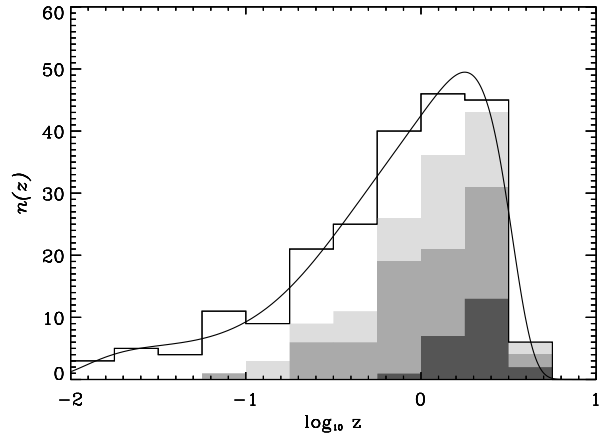}}}
\caption{Histogram of the distribution in $\log_{10} z$ of the radio
  galaxies in the 3CRR, 6CE, 6C* and 7CRS samples. The bin width is
  $\Delta (\log_{10} z) = 0.25$. The white regions correspond to 3CRR
  sources, the light shaded regions to 6CE sources, the intermediate
  shaded regions are the 7CRS sources, and the dark shaded regions the
  6C* sources. The solid line represents the best-fitting function to
  the data -- $n(z) \times \Delta (\log_{10} z)$ -- given by
  Eq.~\ref{eq:Nz} and the coefficients in the text.}
 \label{fig:histlogz}
\end{center}
\end{figure}

In this section we consider how to model the distribution of galaxies
in the $K-z$ diagram.  We start from the combined dataset of radio
galaxies from the 3CRR, 6CE, 6C* and 7CRS samples\footnote{These data
can be obtained on-line at
http://www-astro.physics.ox.ac.uk/$\sim$cjw/kz/kz.html}, and fit their
distribution in $\log_{10} z$ with a function of the form:
\begin{equation}
n (z) = A \exp \bigg\{ -\biggl[ \sum_{i=0}^n a_{i}\, (\log_{10}\, z)^i \biggr]^2 \bigg\},
\label{eq:Nz}
\end{equation}
where $A$ and $a_{i}$ are free parameters.  The histogram of the
distribution in $\log_{10} z$ along with the best-fitting function are
shown in Fig.~\ref{fig:histlogz}. We use a fifth order polynomial as
the argument to the exponential, and find the best-fitting
coefficients to be:
\begin{displaymath}
\begin{array}{cccc}
A  =  197.96; &  a_{0}  =  -0.39;  & a_{2}  =  1.00; & a_{4}  =  1.47;\\    
              &  a_{1}  =  ~~1.17;   & a_{3}  =  1.83; & a_{5}  =  0.38.\\\end{array}
\end{displaymath}
For any given value of redshift $z$, we assume that $n(z)$ of the
sources follow a Gaussian distribution in \kband magnitude about a
mean value $k(z)$, given by the $K-z$ relation (Eq.~\ref{eq:K-z}) at
that redshift, i.e.:
\begin{equation}
\rho (K|z) = \frac{n(z)}{\sigma_{K} \sqrt{2\pi}}\,\,\exp\left\{-\frac{[K- k(z)]^2}{2(\sigma_{K})^2}\right\},
\label{eq:model}
\end{equation}
where $K$ is the aperture- and emission-line corrected \kband
magnitude measured for any given source (as described in Jarvis et
al. 2001a; Willott et al. 2003) and $\sigma_{K}$, the dispersion, is
independent of redshift. This assumption is motivated by the results
of Willott et al. (2003), who evaluated the dispersion about the
best-fitting $K-z$ relation as a function of redshift. They found no
significant increase up to $z = 3$, which is also in agreement with
the results of Jarvis et al. (2001a).

In order to constrain the free-parameter $\sigma_{K}$ we follow the
maximum likelihood formulation of Marshall et al. (1983).  Defining
$S$ as $-2\, {\rm ln}\, \mathcal{L}$, where $\mathcal{L}$ is the
likelihood function, the best-fit parameter is obtained by minimizing
the value of S, which in our case is given by:
\begin{equation}
S = - 2 \sum_{i=1}^{N} \ln[\rho (K_{i}|z_{i})] + 2 \int\int \rho (K|z){\rm
  d}K\,{\rm d}( \log_{10}\,z). 
\label{eq:marshall}
\end{equation}
In the first term the sum is over all $N$ radio galaxies in the
combined sample; the second term is the integral of the model
distribution being tested and should give $\approx 2N$ for good
fits. The upper and lower limits of the integral are: $10.0 \leq K \leq
21.0$ and $-1.3 \leq \log_{10} z \leq 1.0$ (corresponding to $0.05
\leq z \leq 10$).  To find the best-fitting parameter we evaluated $S$
over a range of $0.01 \leq \sigma_{K} \leq 1$.  We found a minimum
value of $S$ for $\sigma_{K} = 0.593 \pm 0.02$\,mag.  This value is in
good quantitative agreement with the results of Willott et al. (2003),
who found a scatter of 0.58\,mag in the $K-z$ relation at all
redshifts up to $z = 3$.

\subsection{Redshift probability distributions from the simulated $K-z$ diagram 
}
\label{sec:pdf}
Adopting the model described in the previous section, such that $\rho
(K|z)\, {\rm d}K \,{\rm d} (\log_{10} z)$ is the expected number of
sources in the differential magnitude element ${\rm d}K$ and in the
differential redshift element ${\rm d} (\log_{10} z)$, and using
Poisson probabilities, we use Monte Carlo simulations to generate a
large number of samples that mimic the combined 3CRR/6CE/6C*/7CRS
dataset.  We combine all the simulated datasets to construct a highly
populated synthetic $K-z$ diagram from which it is possible to extract
photometric redshift probability density functions for any given value
of K-magnitude. We extract these functions from a total of 10000
simulated samples, i.e. from a $K-z$ diagram with $\sim$ 2 million
sources.

The probability density functions $p(z|K)$ can be obtained for any
given source with $10 \leq K \leq 21$ by taking the points along the
horizontal band on the synthetic $K-z$ diagram defined by $[K-\Delta
K/2; K+\Delta K/2$], where $K$ is the \kband magnitude measured for that
source and $\Delta K$ an appropriately small number, and by fitting
the values of the relative frequency of each data point along this
band.  For any such source these values are best-fit by a
$\log_{10}$-normal distribution of the random variable $z$ with
probability density function:
\begin{equation}
p(z|K) = \frac{1}{\ln (10)\, z\, \sqrt{2\pi \sigma^2}} \exp \left\{ -\frac{[\log_{10}(z)- \mu]^2}{2\, \sigma^2}\right\}
\label{eq:pdf_lognormal}
\end{equation}
where $\mu$ and $\sigma$ are the mean and standard deviation of the
distribution for the normal random variable $\log_{10} z$.  The
best-fitting estimate for $z$ is thus defined as:

\begin{equation}
z_{\rm{est}} = 10^{\mu},
\label{eq:mean}
\end{equation}
and the asymmetric 68\% confidence interval as:
\begin{equation}
10^{\mu - \sigma} \leq z_{\rm{est}}  \leq 10^{\mu + \sigma}.
\label{eq:ci}
\end{equation}

\begin{figure*}
{\hbox to \textwidth{ \null\null \epsfxsize=0.3\textwidth
\epsfbox{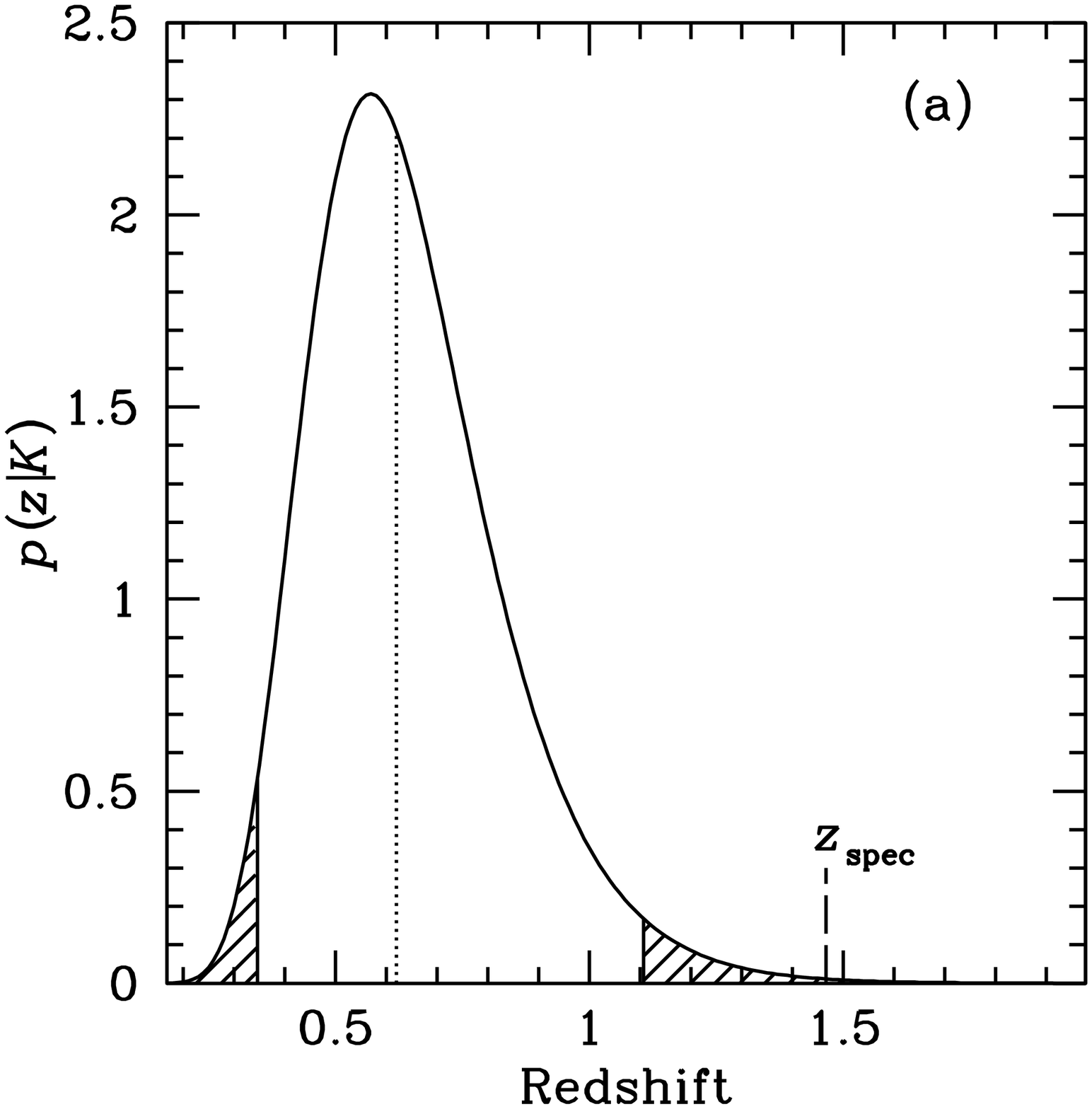}
\epsfxsize=0.3\textwidth
\epsfbox{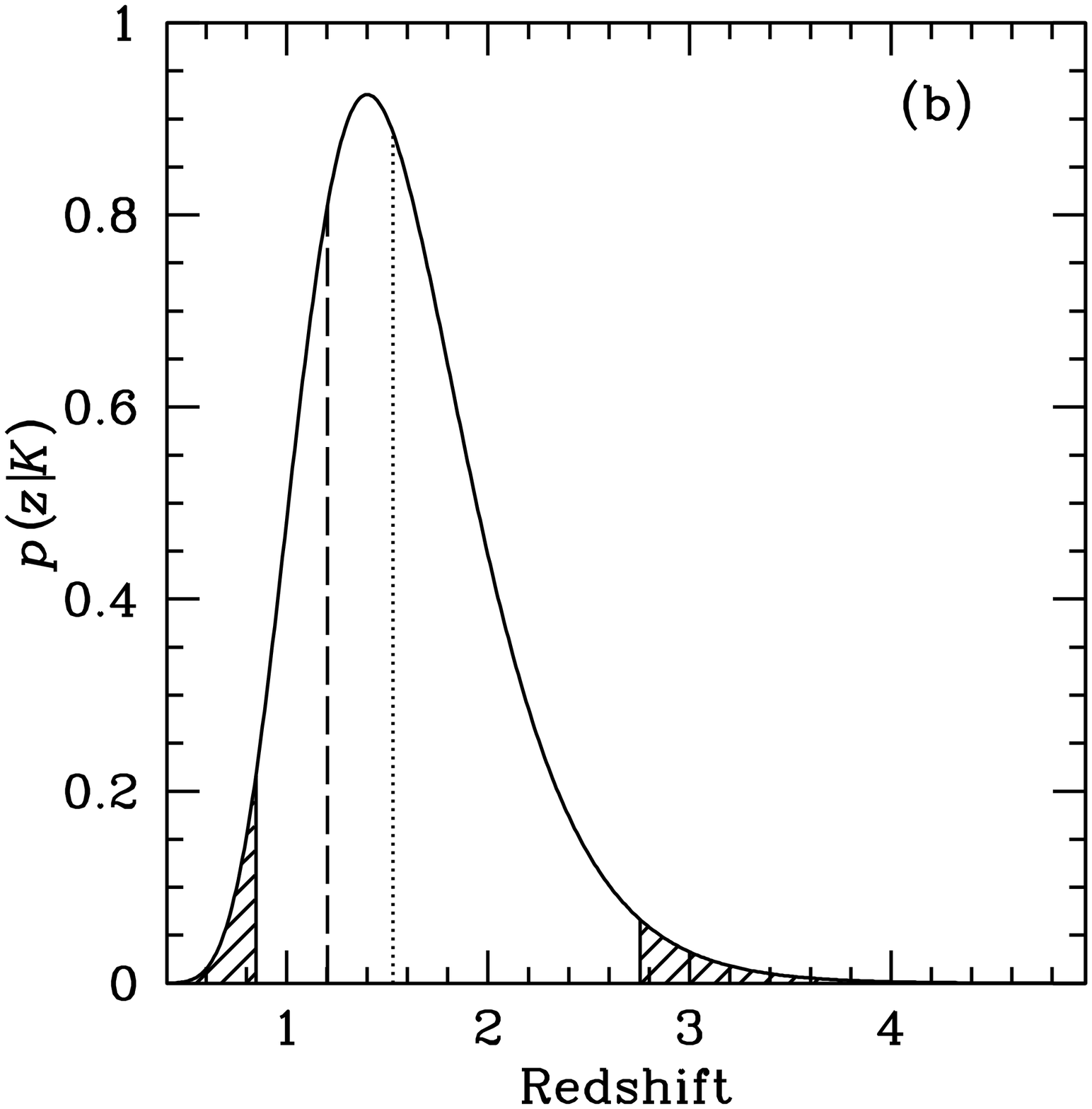} 
\epsfxsize=0.3\textwidth
\epsfbox{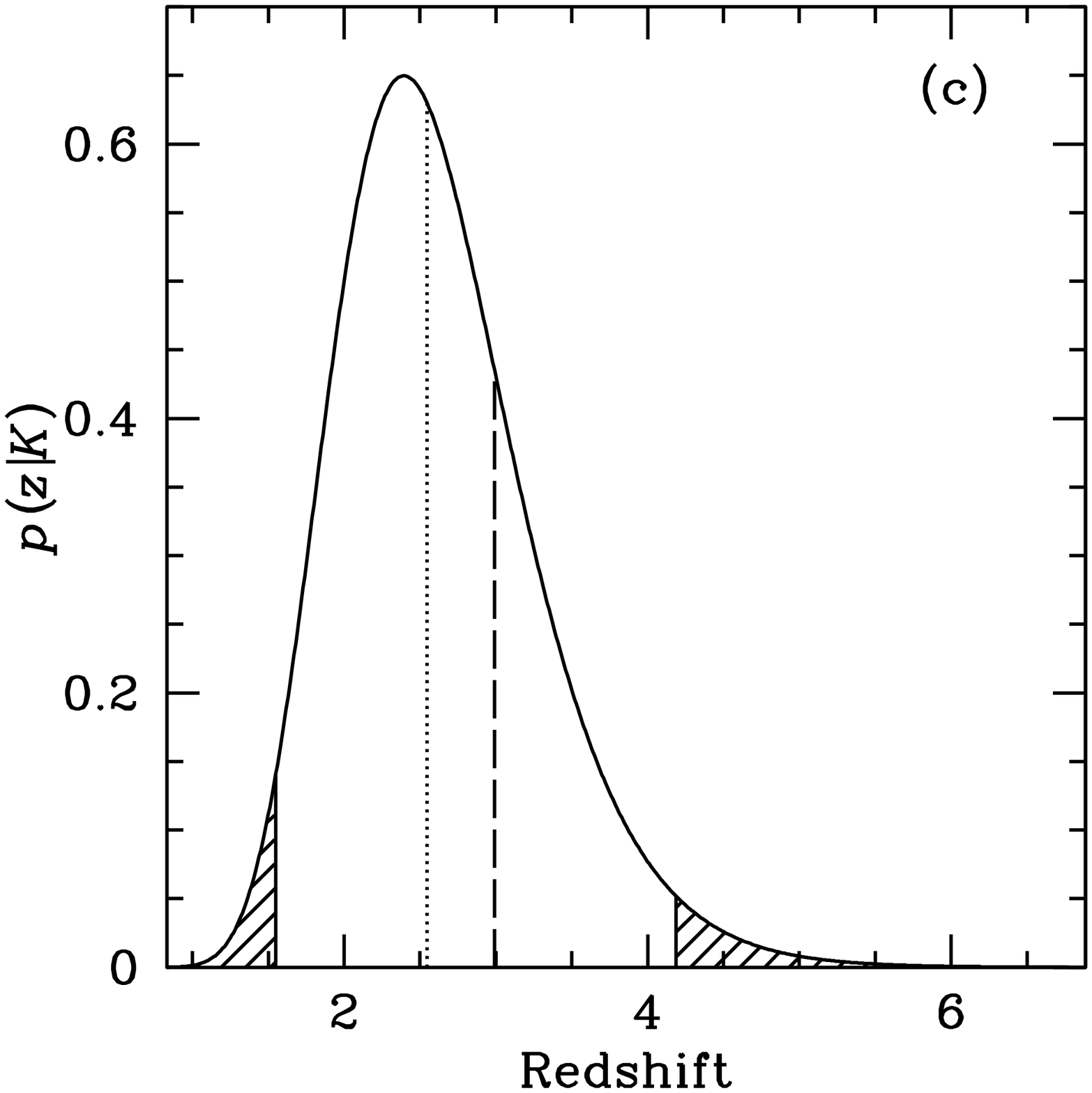} 
}}
{\hbox to \textwidth{ \null\null \epsfxsize=0.3\textwidth
\epsfbox{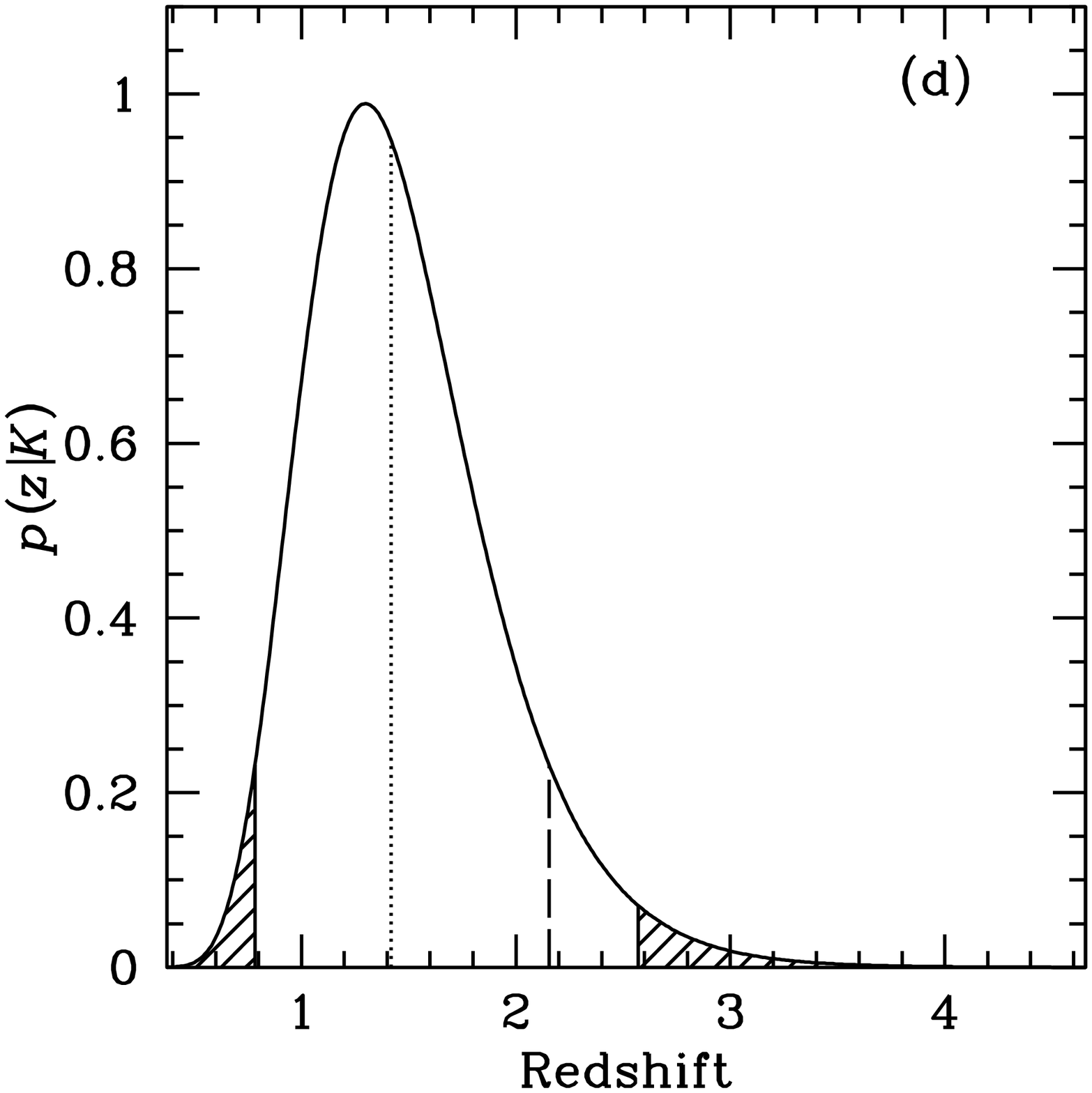}
\epsfxsize=0.3\textwidth
\epsfbox{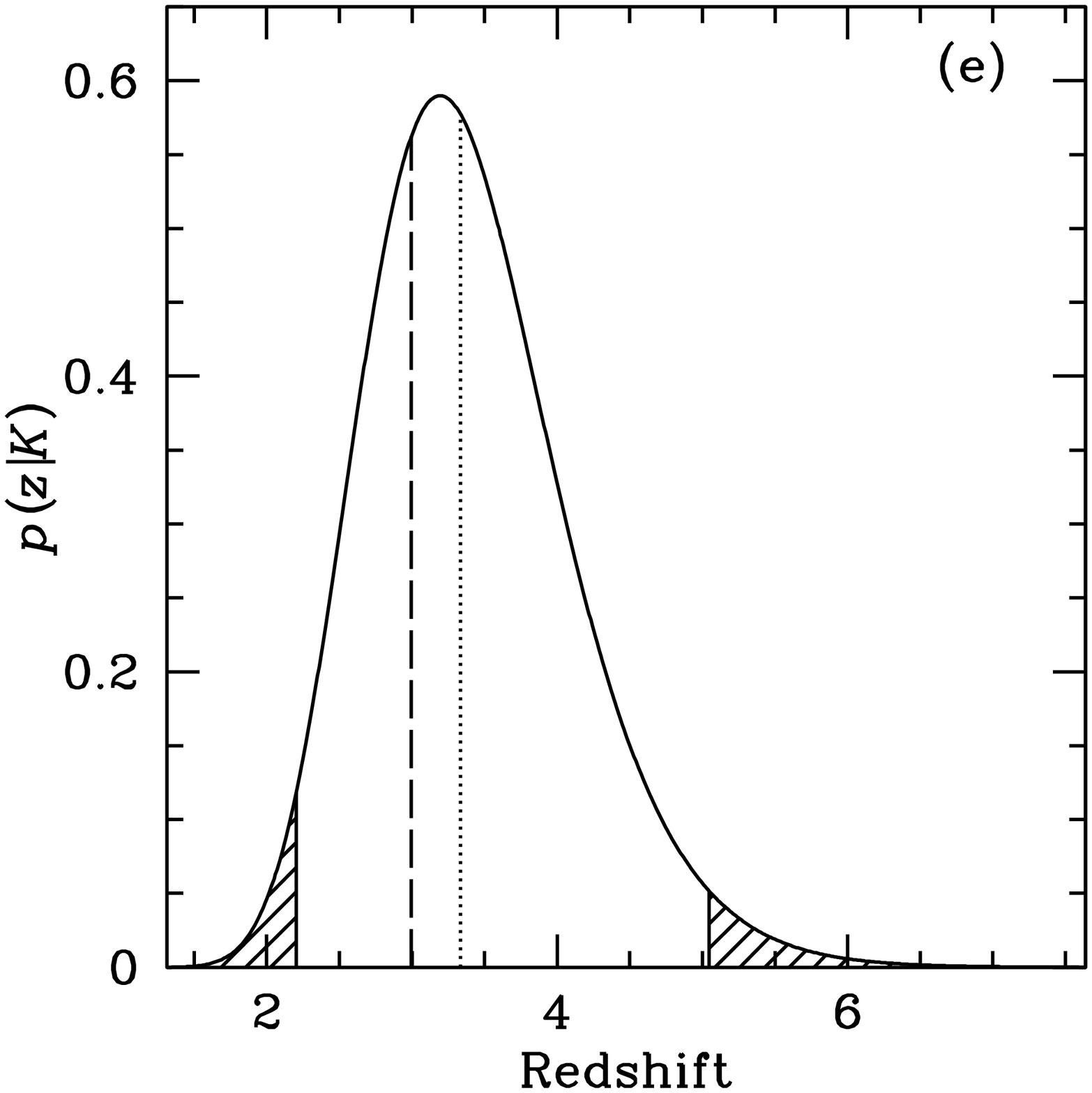} 
\epsfxsize=0.3\textwidth
\epsfbox{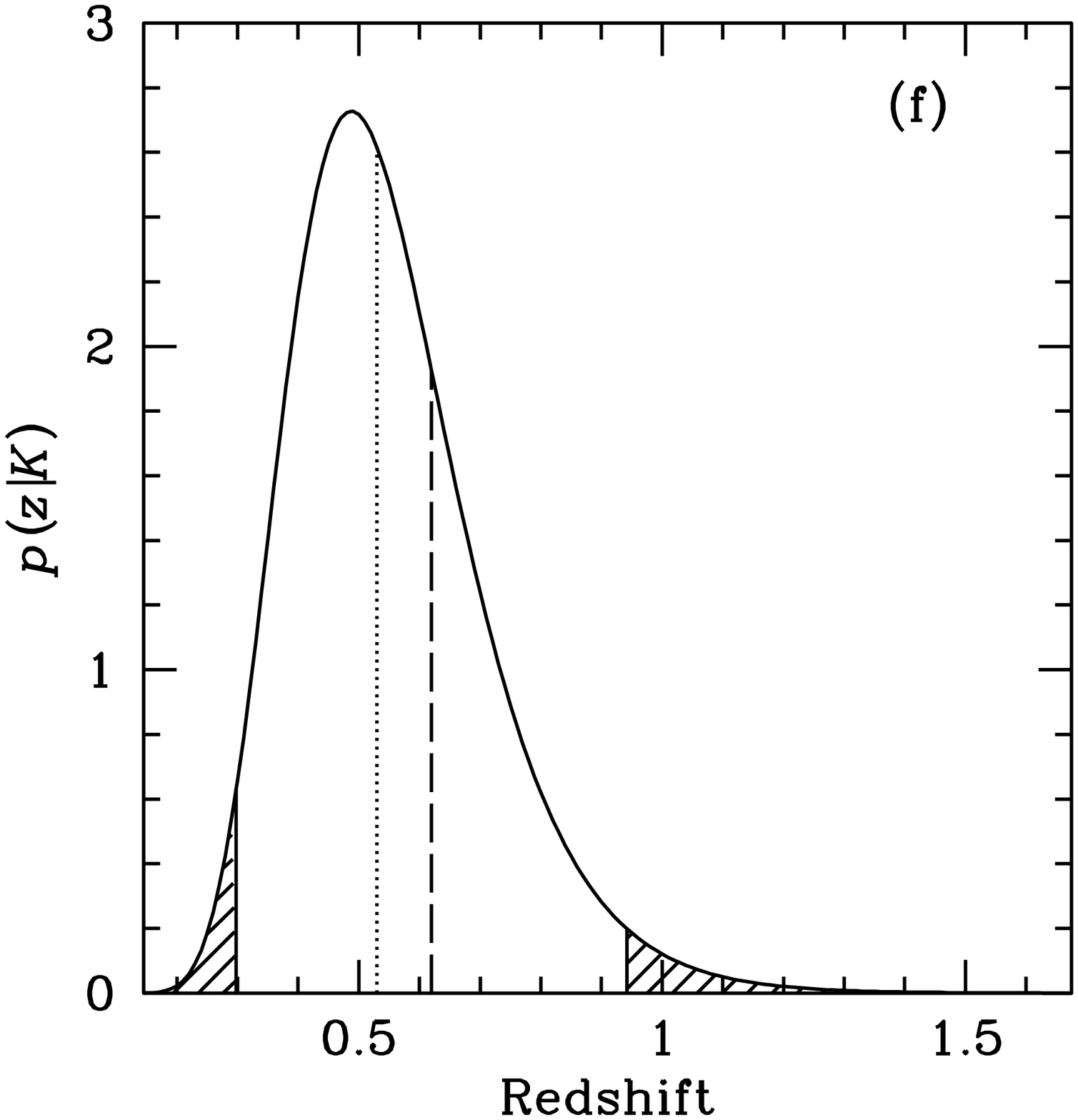} 
}}
{\hbox to \textwidth{ \null\null \epsfxsize=0.3\textwidth
\epsfbox{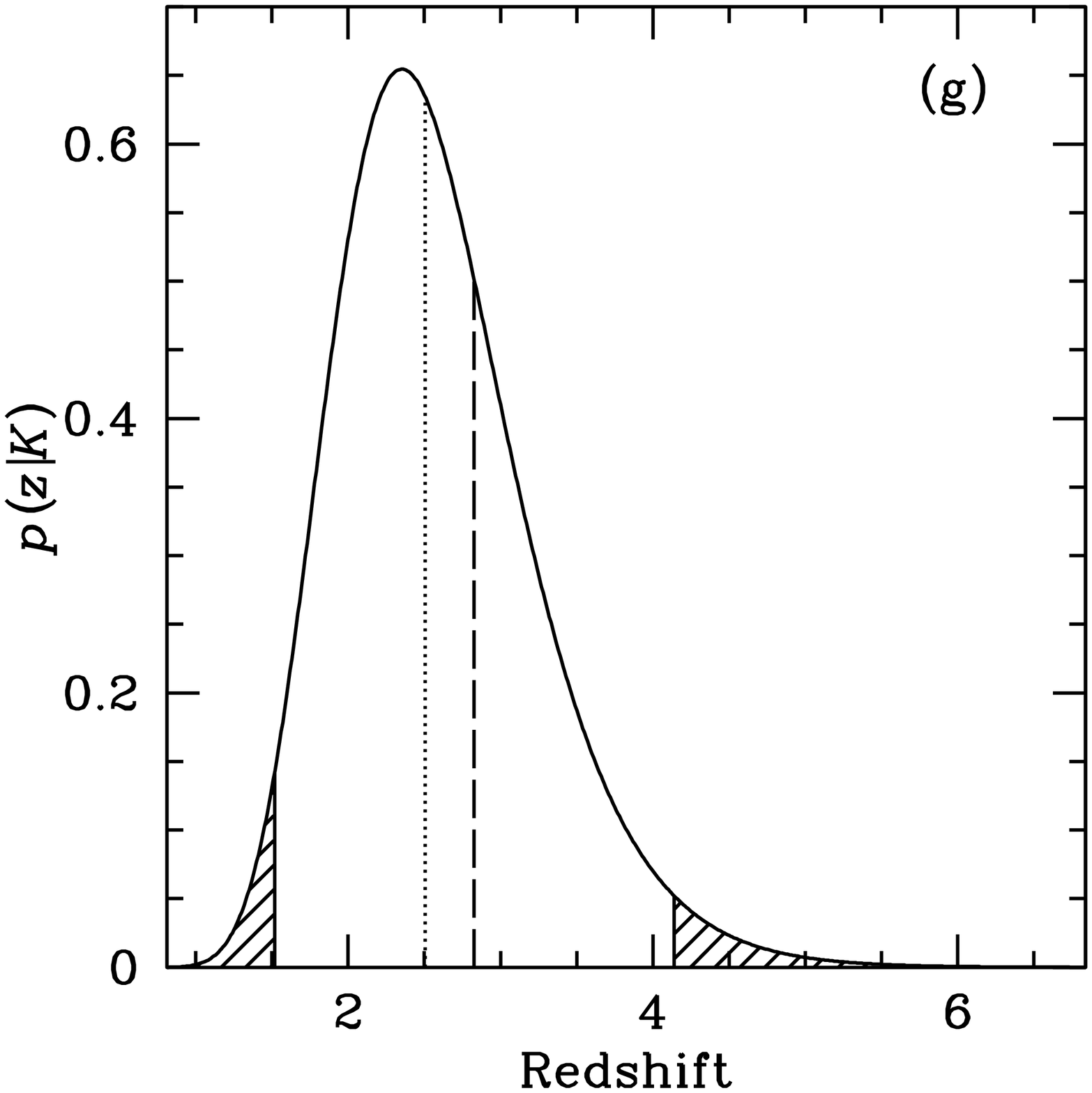}
\epsfxsize=0.3\textwidth
\epsfbox{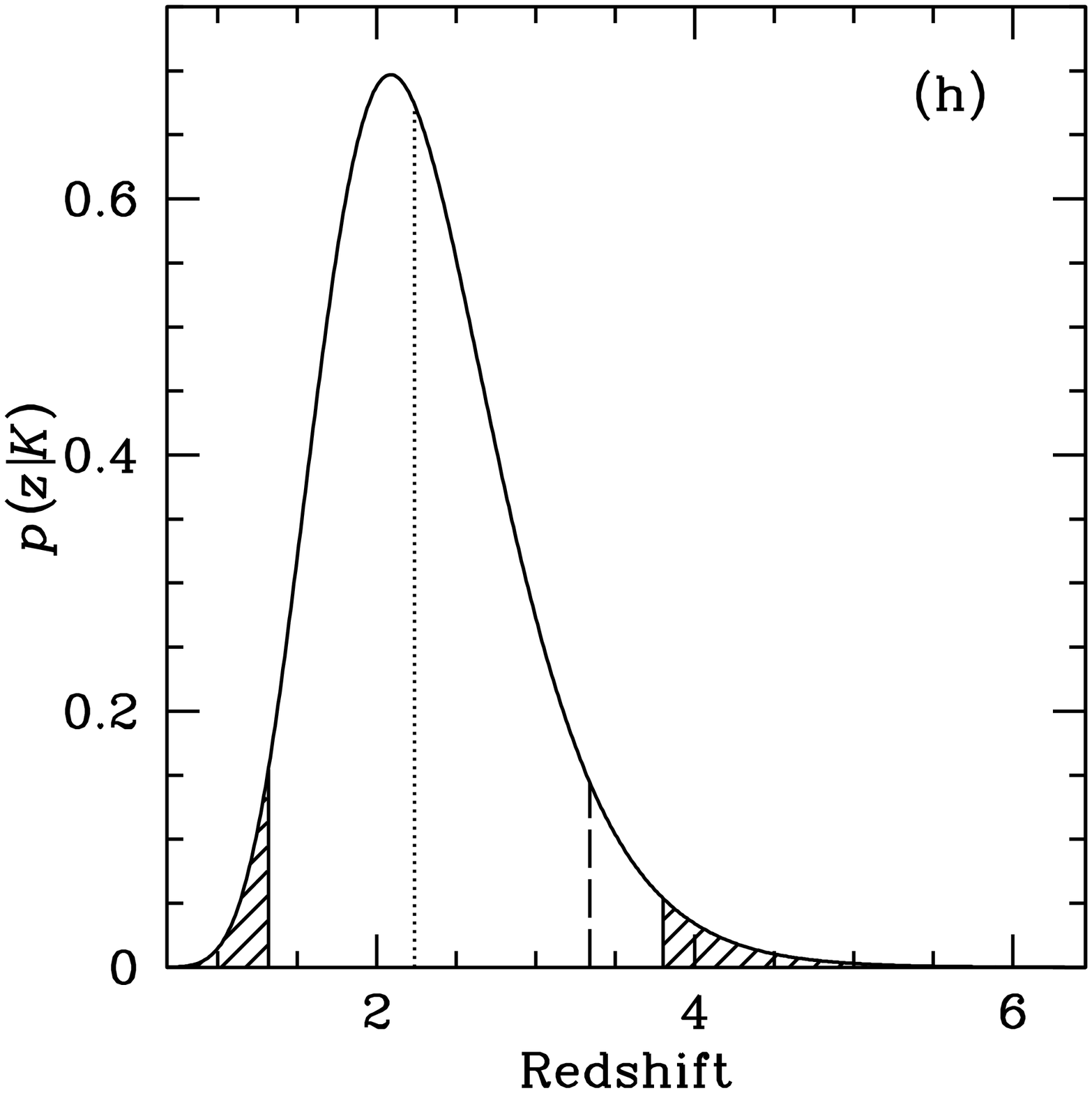} 
\epsfxsize=0.3\textwidth
\epsfbox{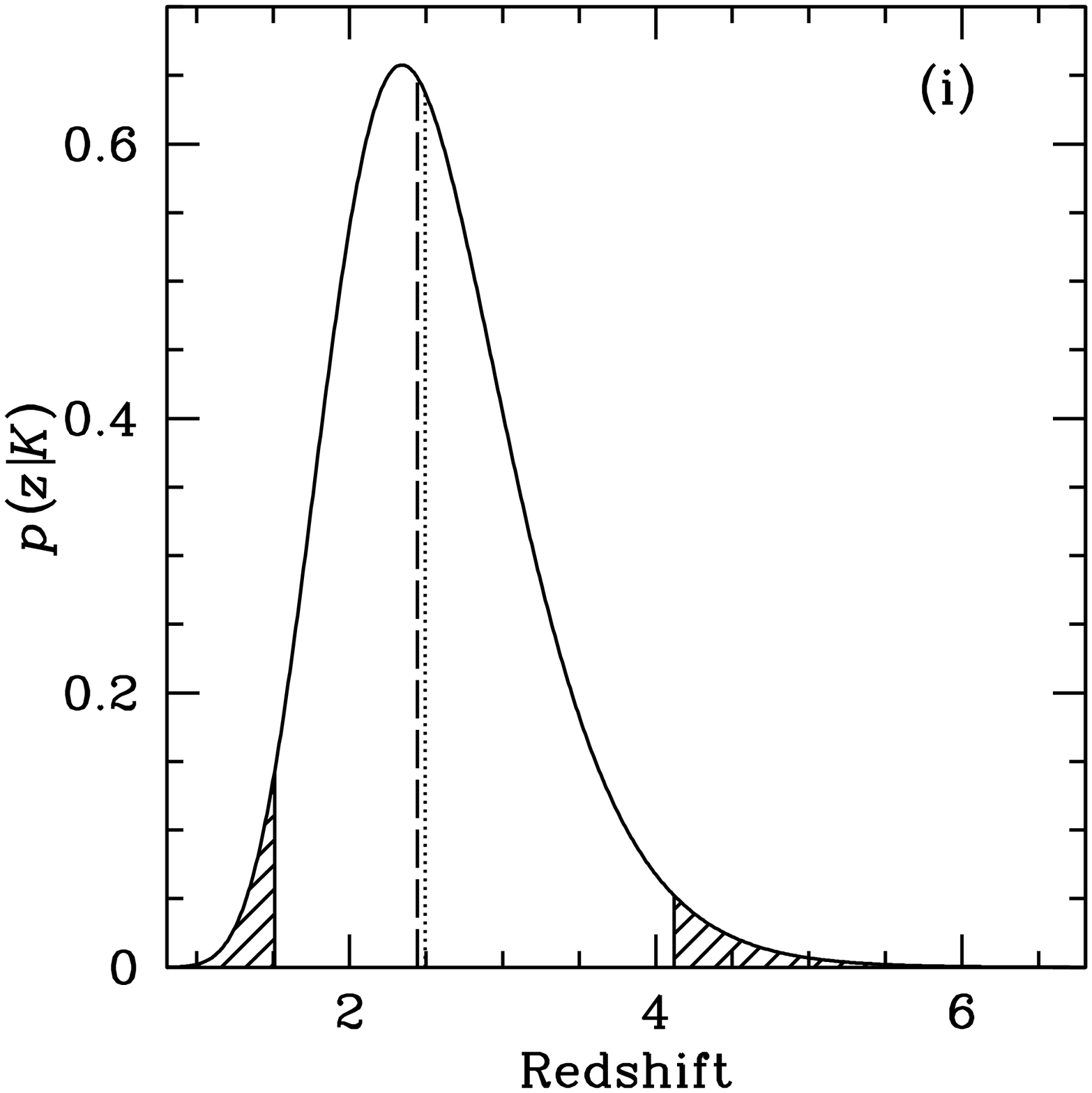} 
}}
{\hbox to \textwidth{ \null\null \epsfxsize=0.3\textwidth
\epsfbox{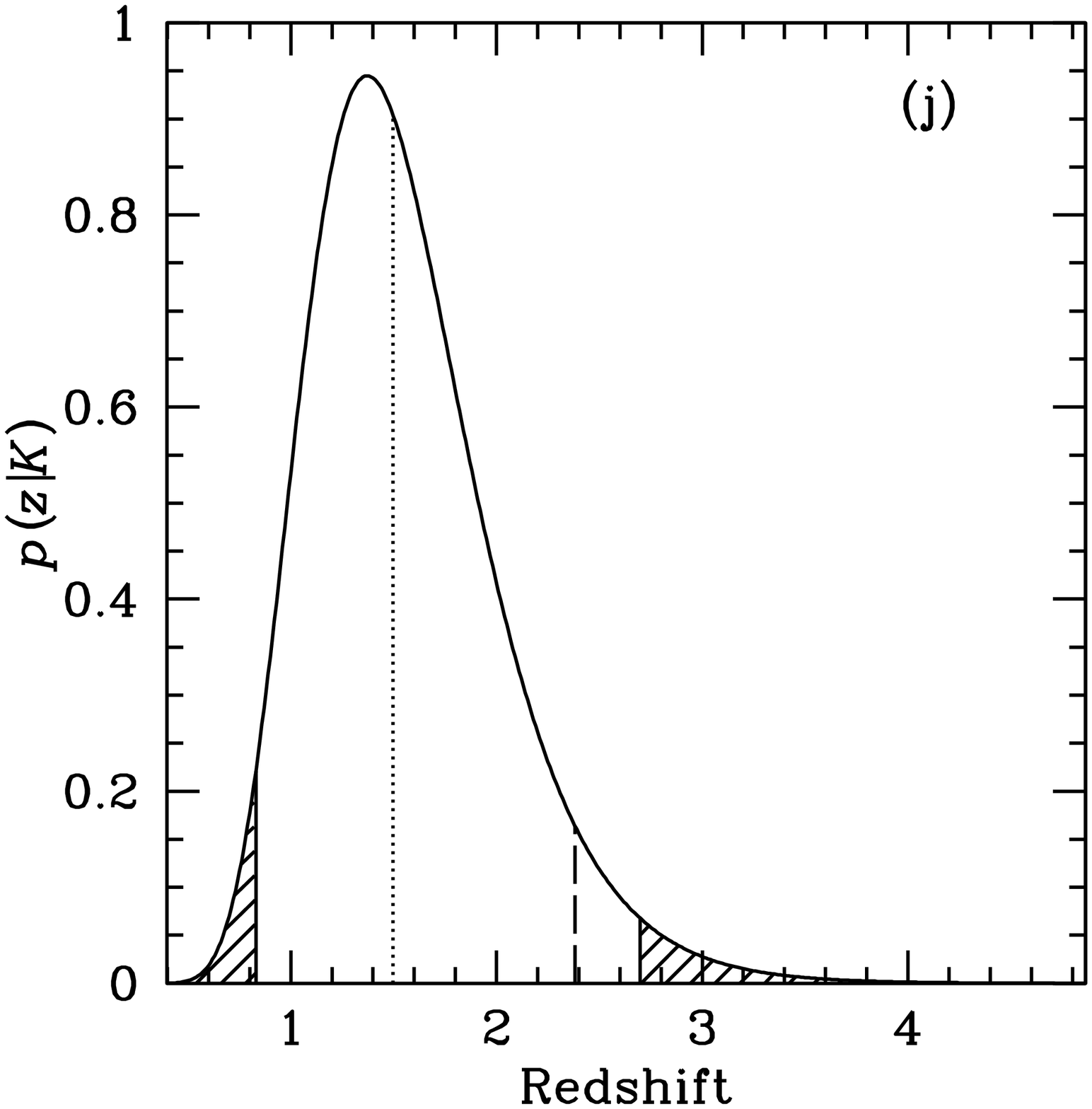}
\epsfxsize=0.3\textwidth
\epsfbox{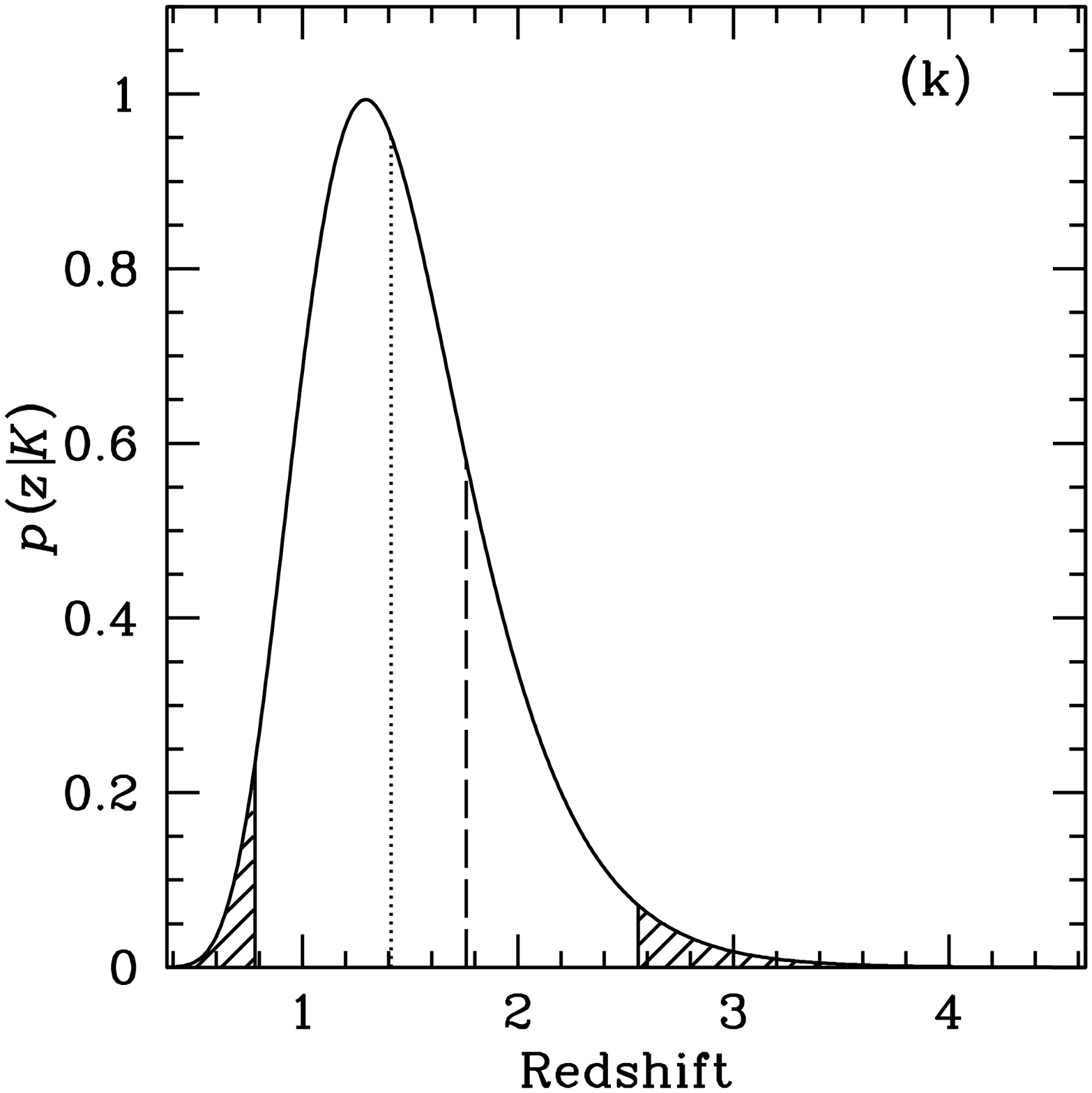} 
\epsfxsize=0.3\textwidth
\epsfbox{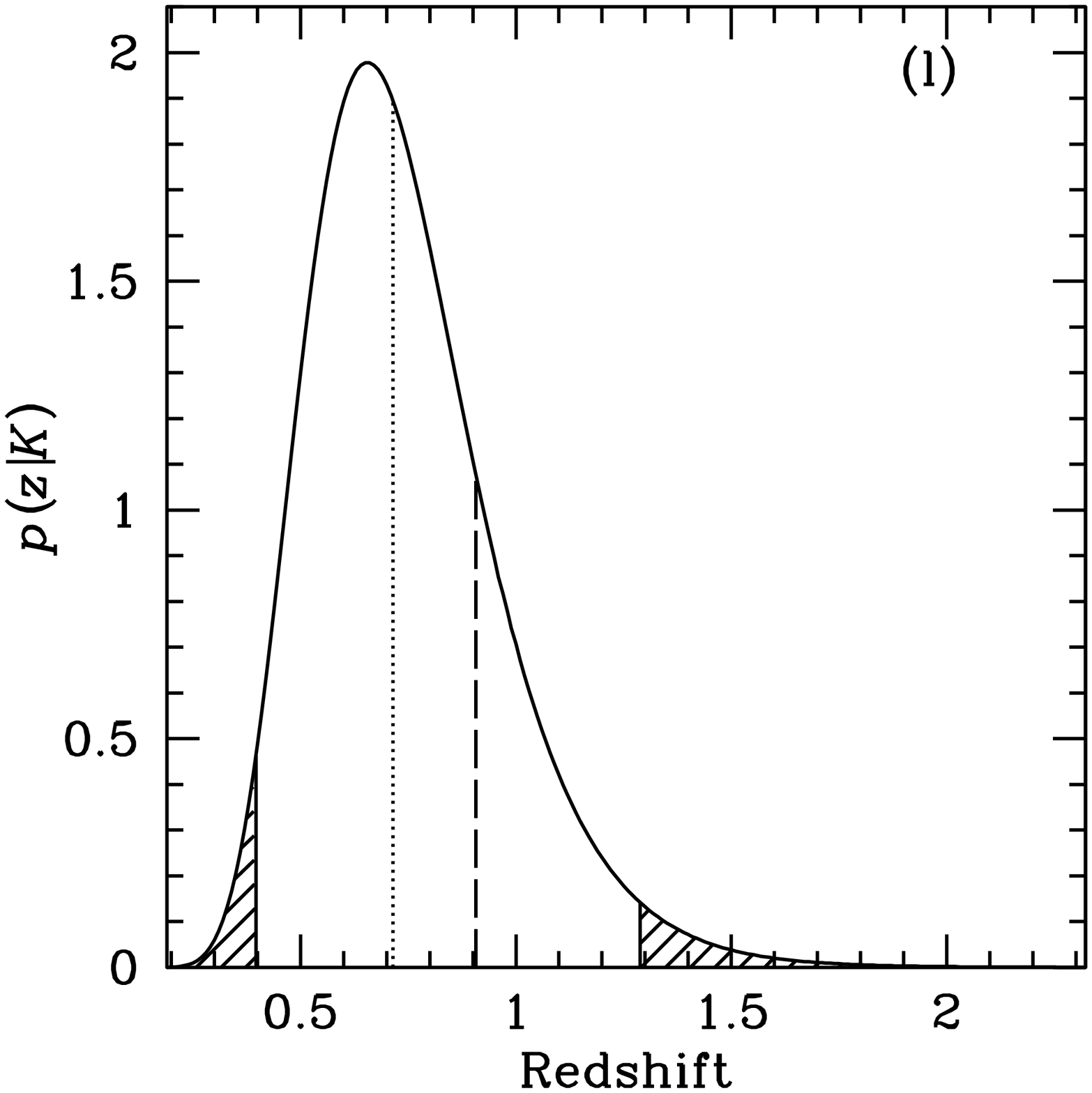} 
}}
\caption{The redshift probability density functions $p(z|K)$ for the
21 (identified) 6C** sources which have spectroscopic redshifts,
normalised such that the area under the curve is unity. For each
graph: the unshaded region corresponds to the 95\% confidence
interval; dotted lines show the location of the redshift estimate;
dashed lines the location of the measured spectroscopic redshift. In
some cases, where the spectroscopic redshift line is difficult to
visualise, we add the label $z_{{\rm spec}}$. (a) {\bf 6C**0714+4616}
(quasar); (b) {\bf 6C**0726+4938}; (c) {\bf 6C**0744+3702}; (d) {\bf
6C**0746+5445}; (e) {\bf 6C**0754+5019}; (f) {\bf 6C**0810+4605}; (g)
{\bf 6C**0824+5344}; (h) {\bf 6C**0832+5443}; (i) {\bf 6C**0834+4129}
(j) {\bf 6C**0854+3500}; (k) {\bf 6C**0856+4313}; (l) {\bf
6C**0903+4251};}
\label{fig:pdfs}
\end{figure*}

\addtocounter{figure}{-1}

\begin{figure*}
{\hbox to \textwidth{ \null\null \epsfxsize=0.3\textwidth
\epsfbox{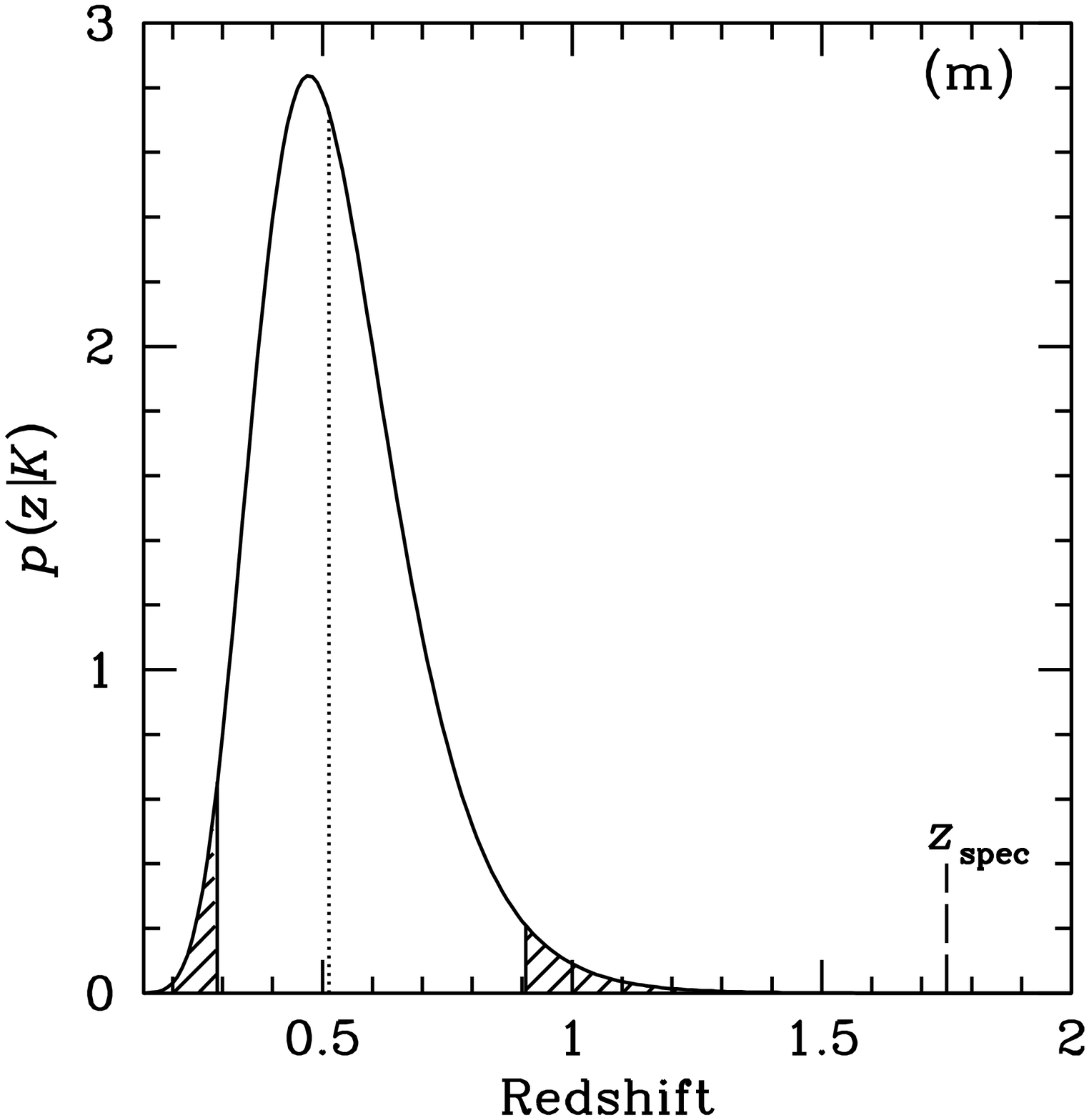}
\epsfxsize=0.3\textwidth
\epsfbox{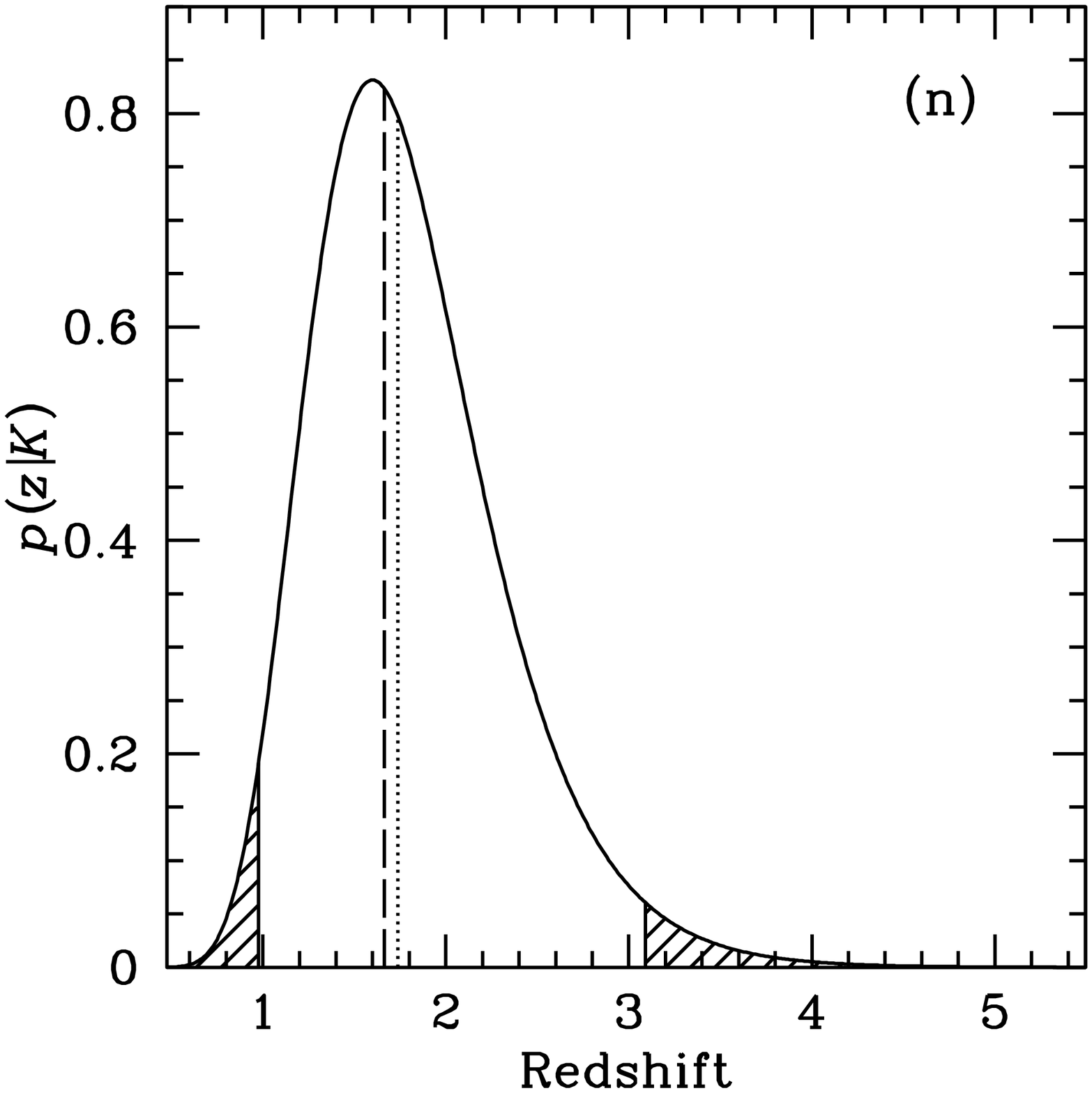} 
\epsfxsize=0.3\textwidth
\epsfbox{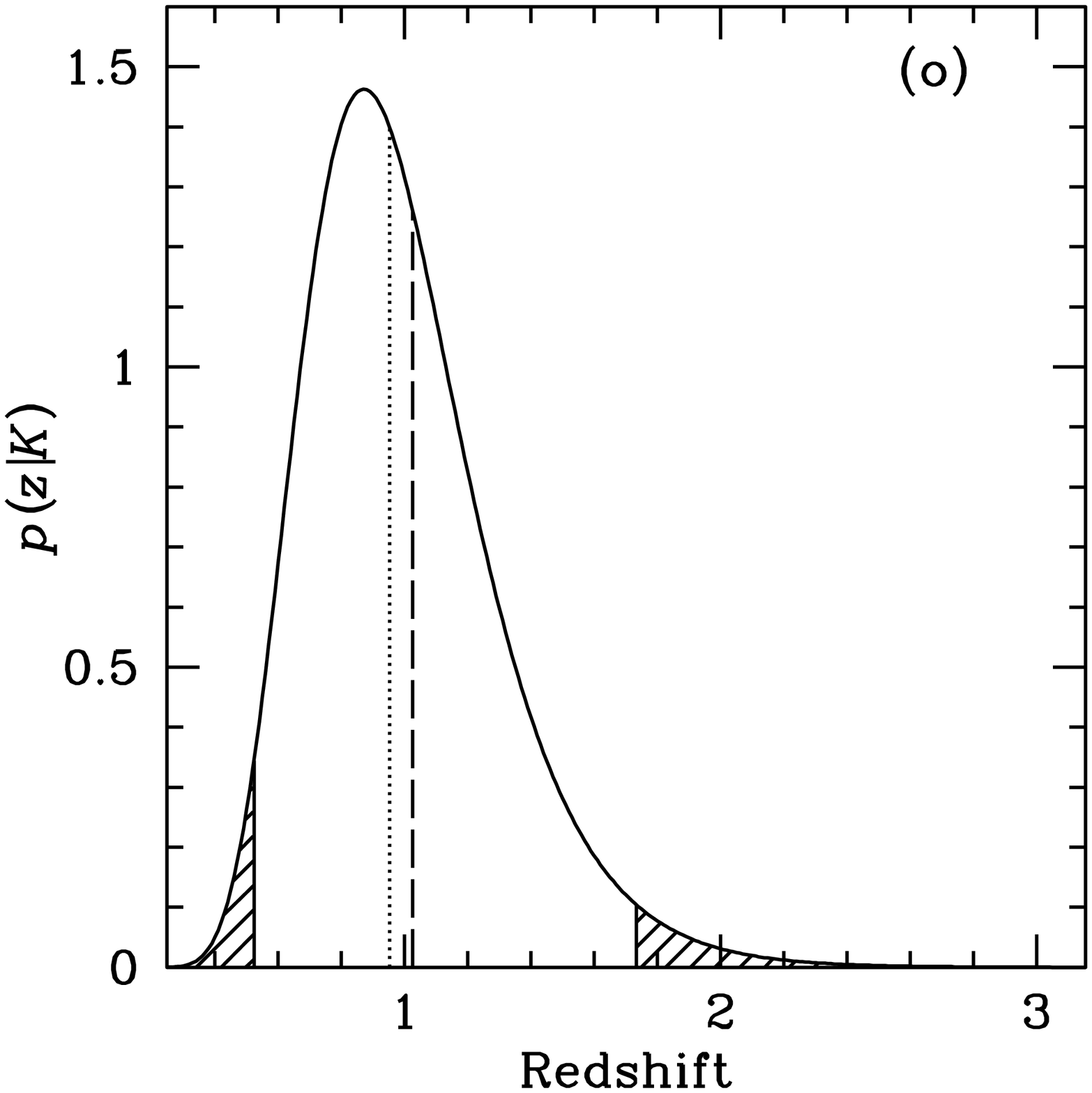} 
}}
{\hbox to \textwidth{ \null\null \epsfxsize=0.3\textwidth
\epsfbox{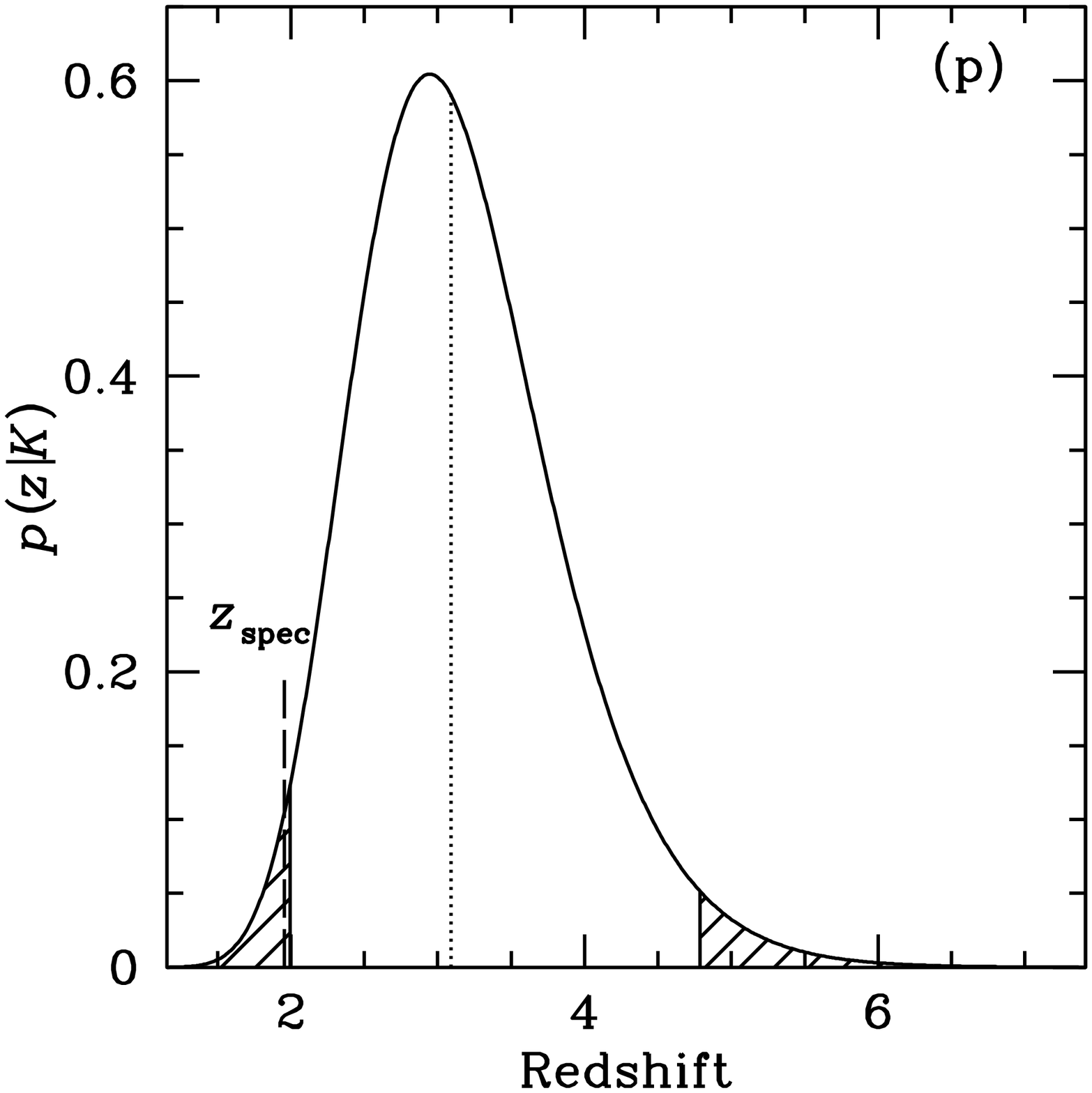}
\epsfxsize=0.3\textwidth
\epsfbox{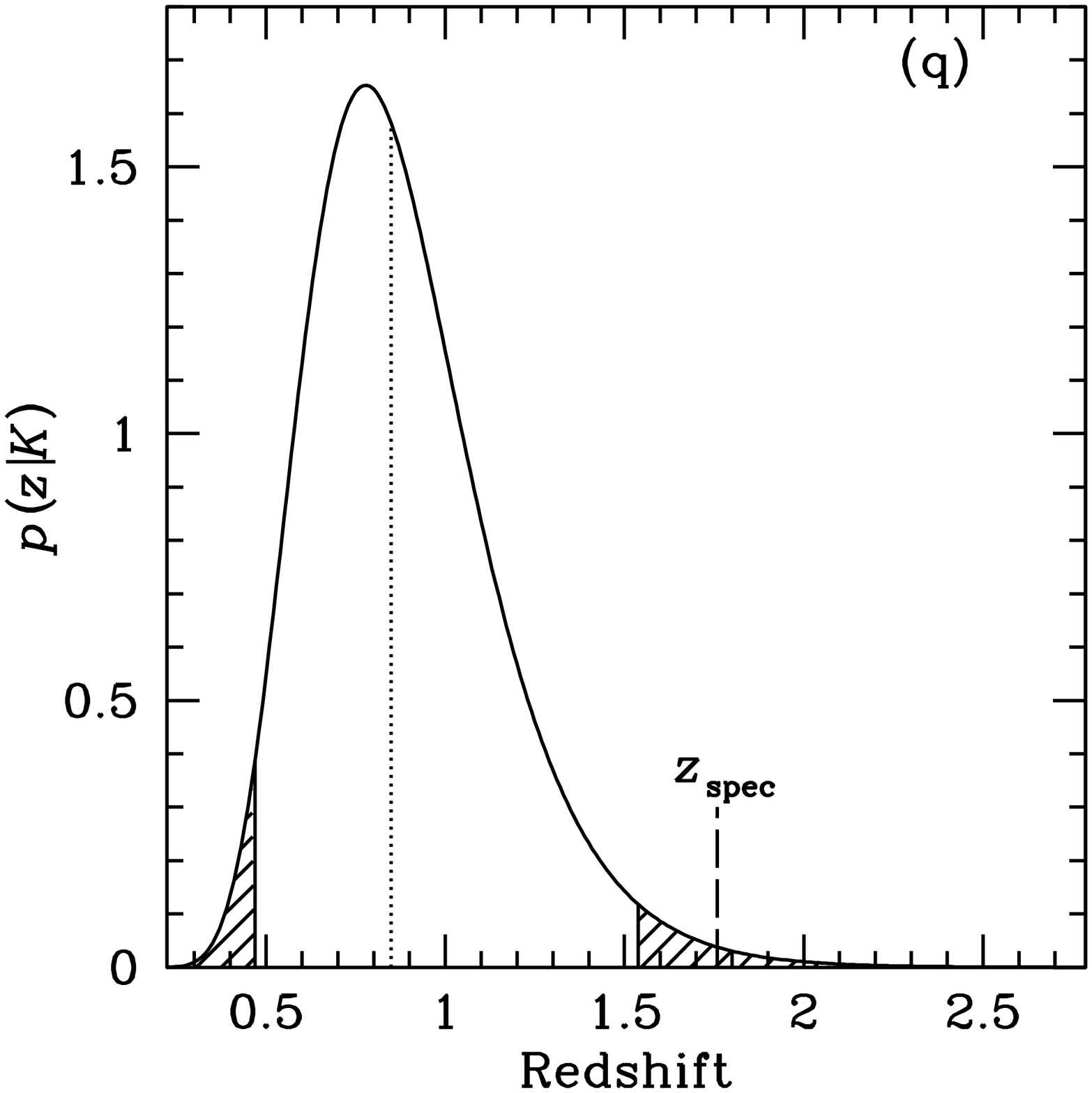} 
\epsfxsize=0.3\textwidth
\epsfbox{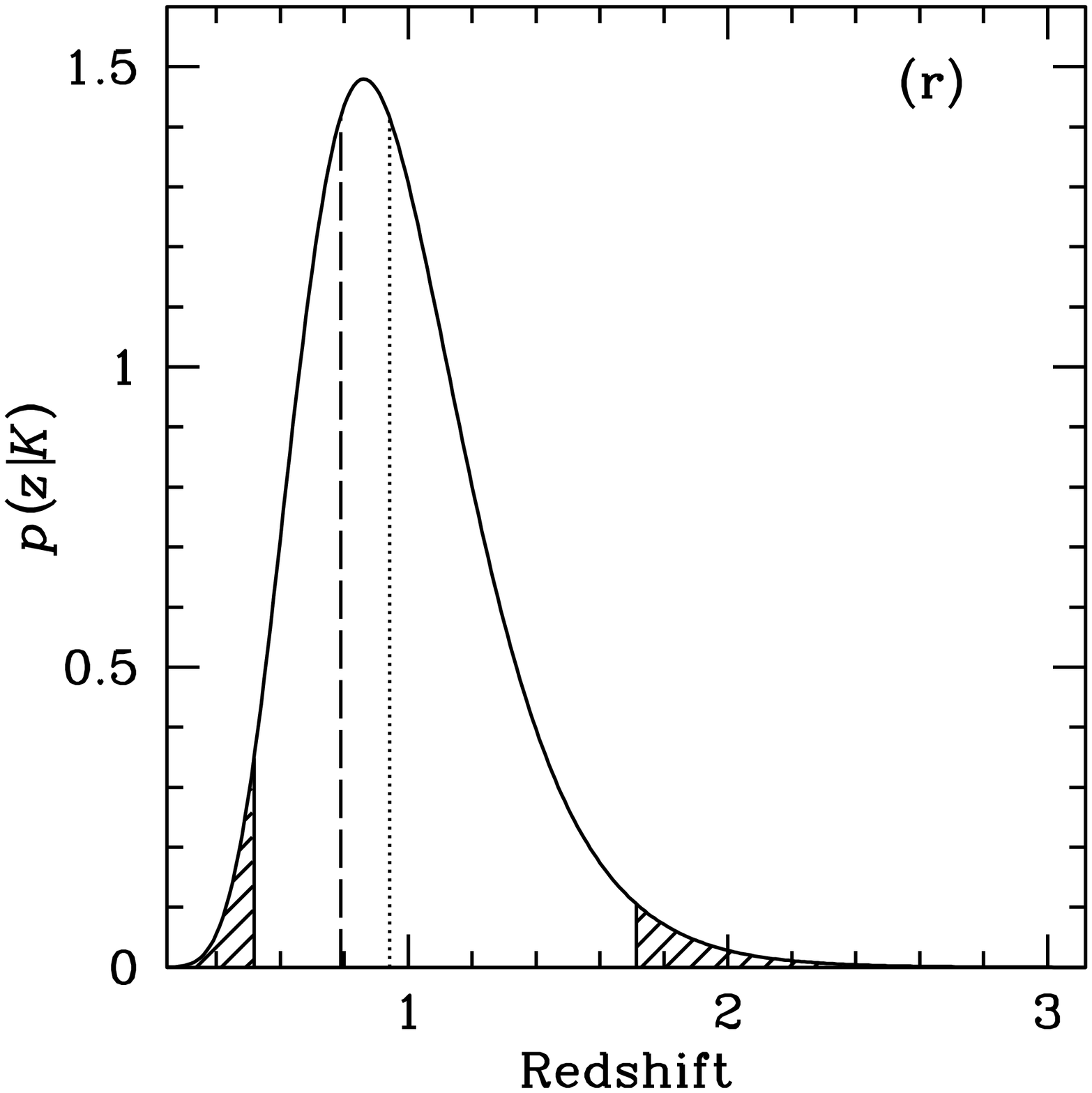} 
}}
{\hbox to \textwidth{ \null\null \epsfxsize=0.3\textwidth
\epsfxsize=0.3\textwidth
\epsfbox{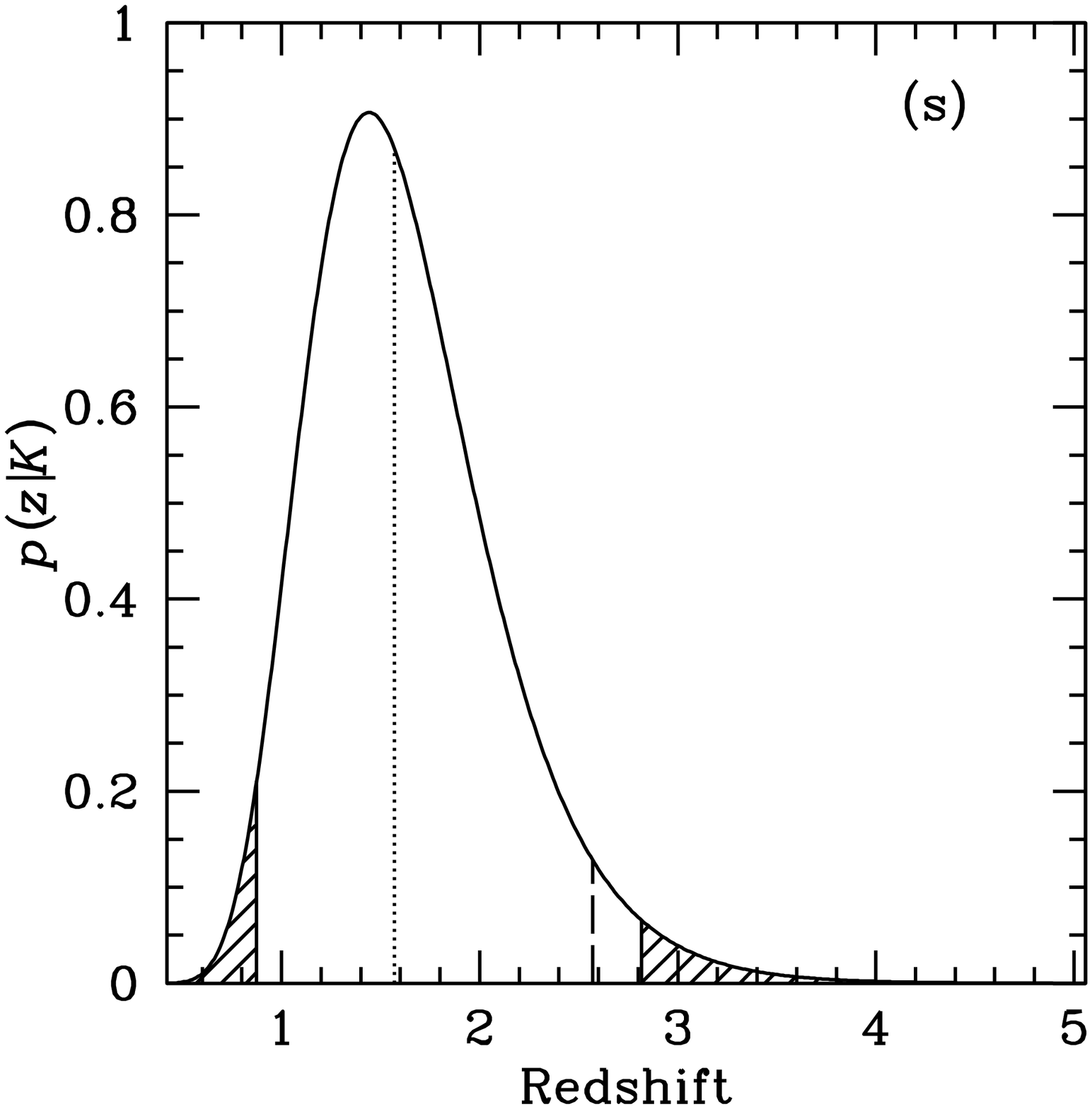} 
\epsfxsize=0.3\textwidth
\epsfbox{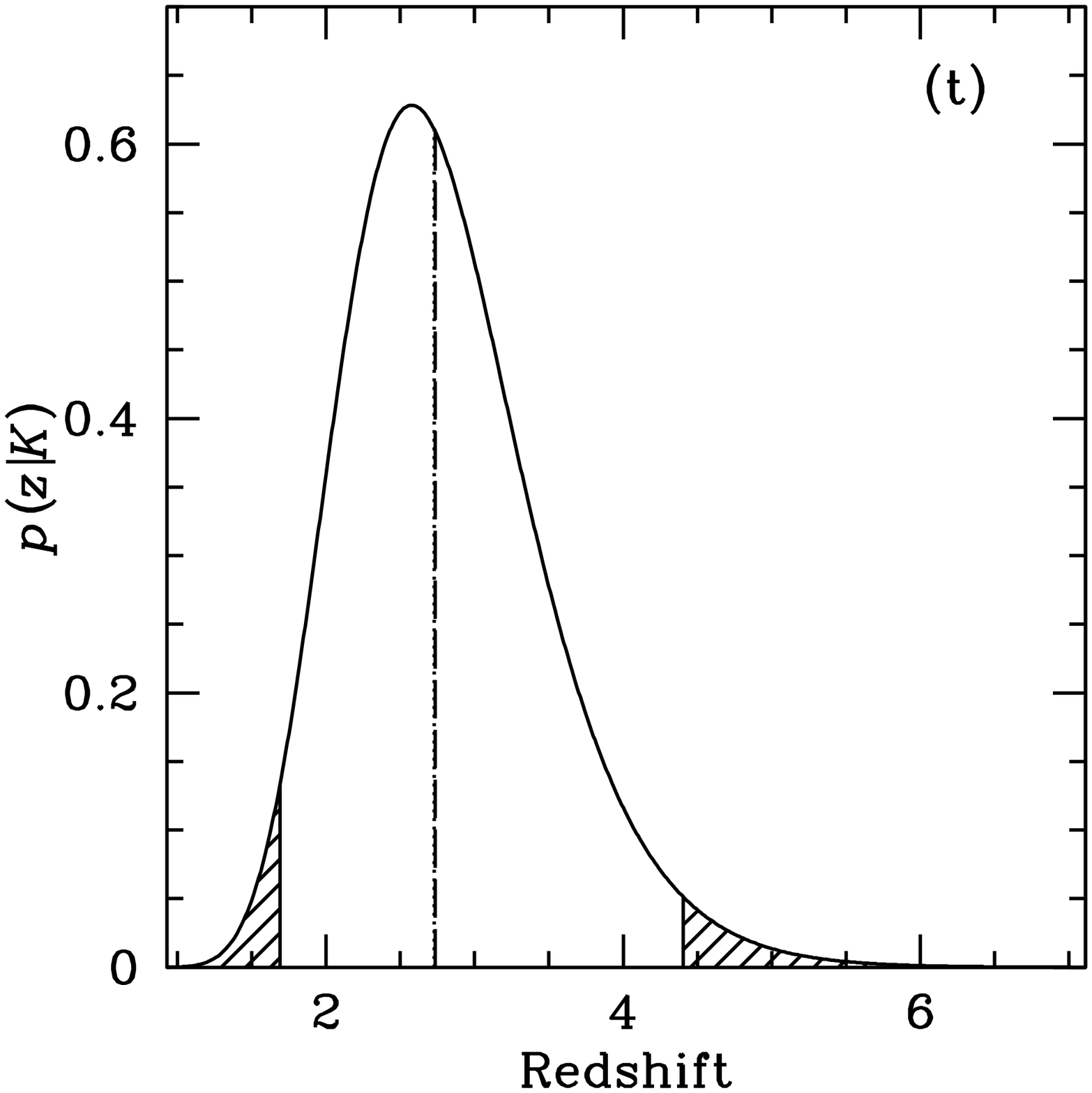}
\epsfxsize=0.3\textwidth
\epsfbox{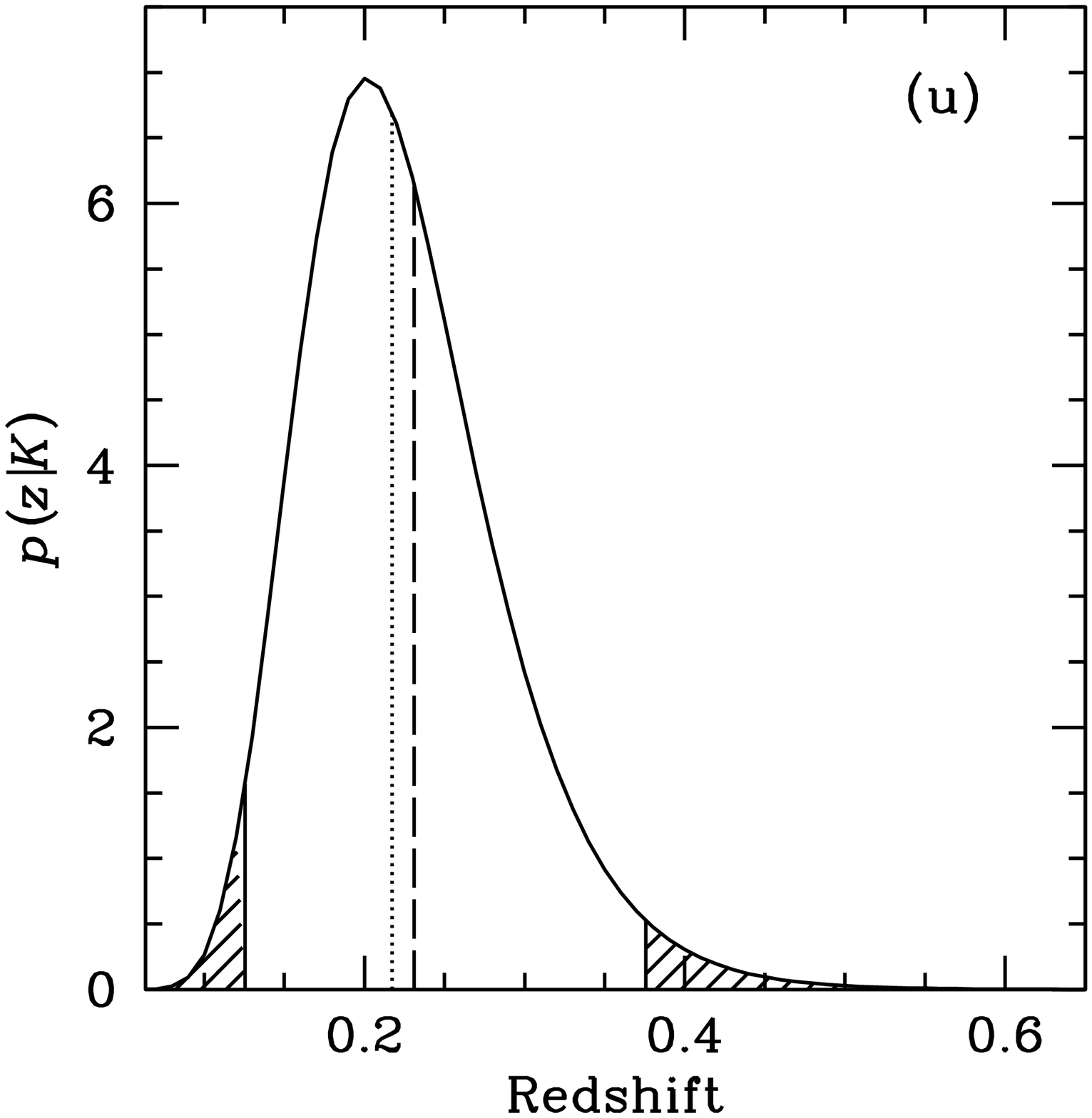} 
}}
\caption{ continued: (m) {\bf 6C**0922+4933} (quasar); (n) {\bf
  6C**0928+4203} (quasar); (o) {\bf 6C**0956+4735}; (p) {\bf
  6C**1009+4327}; (q) {\bf 6C**1036+4721} (quasar); (r) {\bf
  6C**1043+3714}; (s) {\bf 6C**1045+4459}; (t) {\bf 6C**1102+4329};
  (u) {\bf 6C**1132+3209}.}
\label{fig:pdfs}
\end{figure*}

\section{Results}
\label{sec:estimates}

Using \kband photometry from Paper I, we extract photometric redshift
probability density functions for all the identified sources in the
6C** sample\footnote{We exclude just two sources: 6C**0737+5618 and
6C**0935+4348. Both are not identified in \kband down to a 3$\sigma$
limiting magnitude of $K \sim 21$ in an 8-arcsec diameter aperture
(Paper I). Therefore, we do not extract probability density functions
for these sources, as the $K-z$ diagram of radio galaxies is not well
defined for $K > 21$ (see also the caption of
Table~\ref{tab:estimates_median}).}. In
Table~\ref{tab:estimates_median} we quote the values for the
best-fitting redshift estimates $z_{{\rm est}}$ and 68\% confidence
intervals (C.I.), along with the p.d.f parameters $\sigma$ and $\mu$,
and the \kband magnitudes used to extract them. In Fig.~\ref{fig:pdfs}
we present the probability density functions $p(z|K)$ for the 21
identified sources which have spectroscopic redshifts\footnote{This
excludes 6C**0935+4348, which has uncertain \kband identification and
redshift. This source is discussed further in Paper I.} (Paper I and
the references in Table~\ref{tab:estimates_median}).  We note that
there is in general good agreement between the redshift estimates and
the spectroscopic redshifts. The notable exceptions are the quasars,
which have systematically low estimated redshifts because of the
incorrect underlying assumption that the K-band light is dominated by
starlight.

Where possible we have used $K$-magnitudes measured in an 8-arcsec
diameter aperture to extract the probability density functions. For
sources which do not have \kband magnitudes measured in an 8-arcsec
diameter aperture, due to the presence of a nearby object (Paper I),
we have used those measured in 3- or 5-arcsec diameter apertures, and
applied an empirical correction of $-0.21$ and $-0.41$ mag,
respectively. These values have been derived from the median
difference between the small- (3 or 5 arcsec) and large-aperture (8
arcsec) magnitudes (presented in table~5 of Paper I).  This choice of
aperture and these corrections were designed to minimize the effects
of the absence of an aperture correction in our analysis.

We recall that the $K-z$ diagram we use in our modelling is defined in
terms of the aperture- and emission-line corrected \kband
magnitudes. The aperture and emission-line corrections can only be
obtained upon a priori knowledge of redshift and are thereby not
available to us in this analysis. In the remainder of this section we
will discuss how the absence of these corrections affects our redshift
estimates.

\subsection{Aperture correction}

\begin{figure}
\begin{center} 
{\hbox to 0.4\textwidth{ \psfig{figure=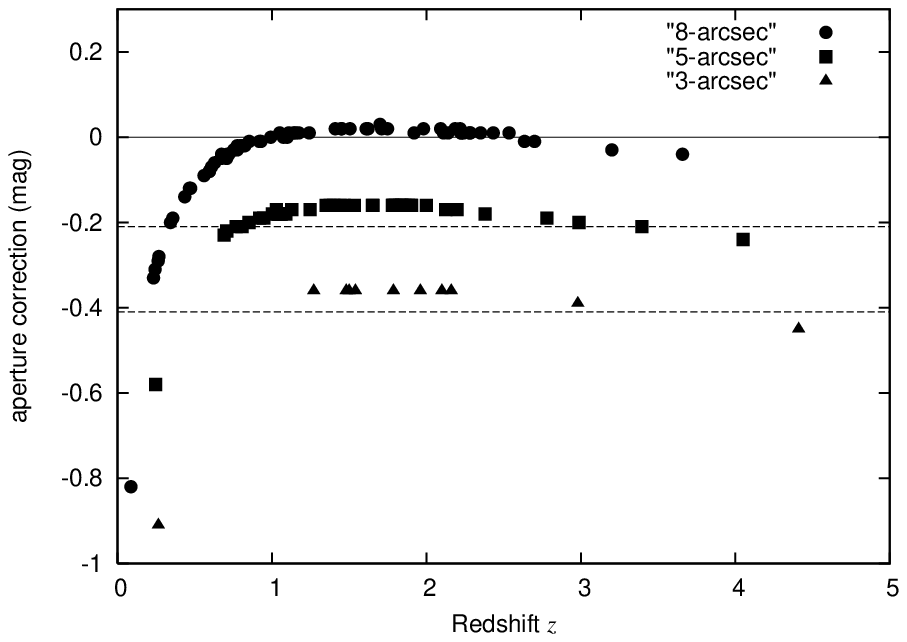,width=0.45\textwidth}}}
\caption{The aperture corrections applied to the \kband magnitudes
measured in 8-arcsec (filled circles), 5-arcsec (filled boxes) and
3-arcsec (filled triangles) diameter apertures for the radio galaxies
in the 3CRR, 6CE, 6C* and 7CRS samples as a function of redshift. The
dashed lines correspond to the empirical corrections applied to the
6C** sources (see Section~ \ref{sec:estimates}) of -0.21 mag (for
magnitudes measured in 5-arcsec apertures) and -0.41 mag (for
magnitudes measured in 3-arcsec apertures).}
 \label{fig:aperture}
\end{center}
\end{figure}

The aperture correction, e.g. as prescribed by Eales et al. (1997),
involves converting the apparent angular size aperture magnitudes to
standard 63.9 kpc metric apertures. This value was chosen because it
corresponds to $\approx 8$ arcsec at $z > 1$ for a $H_{0} =
50$\,km\,s$^{-1}$\,Mpc$^{-1}$, $\Omega_{\rm{M}} = 1$ and
$\Omega_{\Lambda} = 0$ cosmology. As it is immediately apparent from
Fig.~\ref{fig:aperture}, aperture corrections are generally very small
and for $z > 0.6$ not strongly dependent on redshift. For the samples
used in our modelling, at $z > 0.6$ magnitudes measured in 8-arcsec
apertures have aperture corrections which are less than $\pm
0.05$\,mag.  For the magnitudes measured in 5- and 3-arcsec apertures,
the values of the aperture correction are very similar to the
empirical corrections we apply to our data. For these reasons, and for
the high-redshift objects which we are most interested in, we consider
the absence of an aperture correction on a source-to-source basis has
a negligible effect on our redshift estimates.

\subsection{Emission line contribution}
\label{sec:emissionline}

\begin{figure}
\begin{center} 
{\hbox to 0.4\textwidth{ \psfig{figure=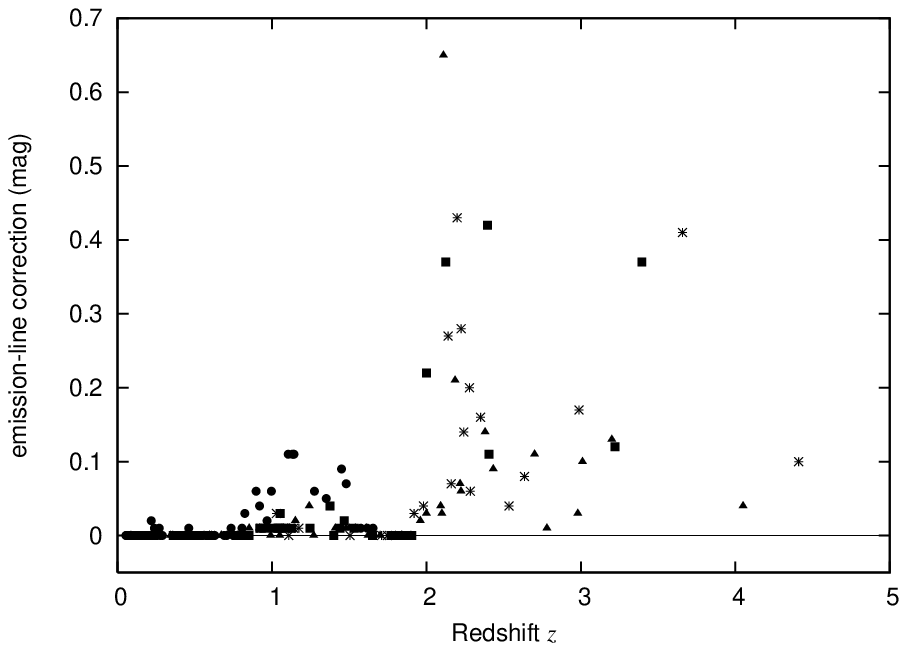,width=0.45\textwidth}}}
\caption{The emission line corrections applied to the \kband
  magnitudes (after aperture correction) of the radio galaxies in the
  3CRR (filled circles), 6CE (filled boxes), 6C* (stars) and 7CRS (filled
  triangles) samples as a function of redshift.}
 \label{fig:emission}
\end{center}
\end{figure}

Emission line contribution to the \kband magnitudes can be significant
at $z > 2$, when the strong H$\alpha$ 6583, [O III] 5007 and [O II]
3727 emission lines are redshifted into the \kband at $1.9 \,\,\ltsim
\,\,z \,\,\ltsim \,\,2.7$, $2.9 \,\,\ltsim \,\,z \,\,\ltsim \,\,3.7$
and $4.2 \,\,\ltsim \,\,z \,\,\ltsim \,\,5.2 $, respectively (Eales \&
Rawlings 1993, 1996; Jarvis et al. 2001a).  The emission line
correction, prescribed by Jarvis et al. (2001a), takes into account
the [O II] emission-line-radio luminosity correlation of Willott
(2001), and the emission-line flux ratios from McCarthy (1993) to
determine the contribution to the \kband magnitude from all the
emission lines at a given redshift. The emission line correction is
therefore dependent both on radio luminosity and on redshift, but more
strongly on redshift. This strong dependency is visible in
Fig.~\ref{fig:emission}, where we plot the emission line correction as
a function of redshift for the radio galaxies in the 3CRR, 6CE, 6C*
and 7CRS samples. However, it can be seen that, apart from one of the
sources which has an emission line correction of 0.65\,mag, all the
sources in this dataset have emission line corrections which are
smaller than the scatter in the $K-z$ diagram. Therefore, we expect
some of the most radio luminous sources in 6C**, which are at $z > 2$
to have an additional systematic uncertainty in their redshift
estimates, due to emission line contamination to their \kband
magnitudes. These will have redshift estimates, based on our method,
which are biased towards lower values. This is a consequence of the
contribution from bright emission lines, which make the objects
brighter (in $K$-band).

\begin{table*}
\begin{center}
\begin{tabular}{lllllllll}
\hline
\mc{1}{c}{Source} & \mc{1}{c}{$K$} & \mc{1}{c}{$\mu$} &
\mc{1}{c}{$\sigma$} & \mc{1}{c}{$z_{\rm{est}}$} & \mc{1}{c}{68\% C.I.}
& \mc{1}{c}{$z_{\rm{spectr}}$} & \mc{2}{c}{$\log_{10}L_{151}$}\\
\mc{1}{c}{} & \mc{1}{c}{} & \mc{1}{c}{} & \mc{1}{c}{} & \mc{1}{c}{} &
\mc{1}{c}{} & \mc{1}{c}{} & \mc{1}{c}{{\tiny est.}} &
\mc{1}{c}{{\tiny spec}.}\\
\hline
6C**0714+4616 & 16.316 & -0.208 & 0.126 & 0.619 & [0.463, 0.827] &
1.466 & ~~26.38 & 27.34 \\
6C**0717+5121 & 17.575 &  0.061 & 0.130 & 1.150 & [0.851, 1.553] & & ~~26.90\\
6C**0726+4938 & 18.168 &  0.184 & 0.128 & 1.529 & [1.139, 2.053] &
1.203? & ~~26.94 & 26.67\\
6C**0737+5618 & $\scriptstyle>$21.1&        &       & $\scriptstyle>$ 4.0&  & &
$\scriptstyle >$28.18\\
6C**0744+3702 & 19.440 &  0.406 & 0.108 & 2.549 & [1.986, 3.273] & 2.992 & ~~27.60 & 27.78\\
\\
6C**0746+5445 & 18.011 &  0.152 & 0.129 & 1.421 & [1.056, 1.911] & 2.156 & ~~26.74 & 27.18\\
6C**0754+4640 & 19.758 &  0.449 & 0.102 & 2.810 & [2.222, 3.555] & & ~~27.60\\
6C**0754+5019 & 20.416 &  0.523 & 0.090 & 3.331 & [2.709, 4.097] & 2.996 & ~~27.97 & 27.85\\
6C**0801+4903 & 19.645 &  0.435 & 0.104 & 2.722 & [2.143, 3.459] & & ~~27.79\\
6C**0810+4605 & 15.993 & -0.276 & 0.125 & 0.529 & [0.397, 0.706] & 0.620 & ~~26.96 & 27.12\\
\\			 		 
6C**0813+3725 & 18.798 &  0.305 & 0.120 & 2.016 & [1.530, 2.658] & & ~~27.18\\
6C**0824+5344 & 19.392 &  0.399 & 0.109 & 2.507 & [1.950, 3.223] & 2.824 & ~~27.58 & 27.71\\
6C**0829+3902 & 19.200 &  0.371 & 0.113 & 2.350 & [1.813, 3.045] & & ~~27.32\\
6C**0832+4420 & 18.915 &  0.324 & 0.118 & 2.110 & [1.607, 2.771] & & ~~27.20\\
6C**0832+5443 & 19.070 &  0.350 & 0.115 & 2.240 & [1.717, 2.922] & 3.341 & ~~27.27 & 27.69\\
\\			    	 
6C**0834+4129 & 19.378 &  0.397 & 0.109 & 2.496 & [1.941, 3.211] & 2.442 & ~~27.29 & 27.27\\
6C**0848+4803 & 17.828 &  0.114 & 0.130 & 1.300 & [0.964, 1.754] & & ~~26.85\\
6C**0848+4927 & 18.222 &  0.195 & 0.127 & 1.567 & [1.168, 2.102] & & ~~27.09\\
6C**0849+4658 & 17.319 &  0.0056& 0.130 & 1.013 & [0.751, 1.366] & & ~~27.18\\
6C**0854+3500 & 18.121 &  0.175 & 0.128 & 1.496 & [1.113, 2.011] & 2.382 & ~~27.02 & 27.52\\
\\			    	 
6C**0855+4428 & 18.272 &  0.205 & 0.127 & 1.604 & [1.197, 2.150] & & ~~27.12\\
6C**0856+4313 & 17.999 &  0.150 & 0.129 & 1.413 & [1.050, 1.901] & 1.761 & ~~26.76 & 27.00\\
6C**0902+3827 & 19.077 &  0.351 & 0.115 & 2.246 & [1.723, 2.927] & & ~~27.71\\
6C**0903+4251 & 16.615 & -0.146 & 0.128 & 0.715 & [0.532, 0.961] & 0.907 & ~~26.78 & 27.03\\
6C**0909+4317 & 18.635 &  0.274 & 0.122 & 1.881 & [1.420, 2.493] & & ~~27.83\\
\\	     		    	 
6C**0912+3913 & 17.595 &  0.065 & 0.131 & 1.162 & [0.860, 1.570] & & ~~26.53\\
6C**0920+5308 & 14.526 & -0.576 & 0.121 & 0.265 & [0.201, 0.351] & & ~~24.99\\
6C**0922+4216 & 15.928 & -0.290 & 0.124 & 0.512 & [0.385, 0.682] & 1.750 & ~~26.35 & 27.67\\
6C**0924+4933 & 14.955 & -0.490 & 0.122 & 0.324 & [0.244, 0.429] & & ~~25.41\\
6C**0925+4155 & 20.069 &  0.486 & 0.098 & 3.060 & [2.444, 3.831] & & ~~27.77\\
\\	     				 
6C**0928+4203 & 18.448 &  0.240 & 0.125 & 1.736 & [1.302, 2.315] & 1.664 & ~~27.62 & 27.57\\
6C**0928+5557 & 17.072 & -0.048 & 0.130 & 0.896 & [0.664, 1.208] & & ~~26.28\\
6C**0930+4856 & 18.903 &  0.322 & 0.118 & 2.100 & [1.599, 2.759] & & ~~27.24\\
6C**0935+4348 &$\scriptstyle >$20.9 &        &       &$\scriptstyle>$ 4.0&                 &
2.321? &  $\scriptstyle >$28.38 & 27.75\\
6C**0935+5548 & 18.325 &  0.216 & 0.126 & 1.644 & [1.229, 2.199] & & ~~27.11\\
\\
6C**0938+3801 &  18.132 &  0.177 & 0.128 & 1.504 & [1.119, 2.021] & & ~~27.12\\
6C**0943+4034 &  17.592 &  0.064 & 0.131 & 1.160 & [0.858, 1.567] & & ~~26.79\\
6C**0944+3946 &  19.088 &  0.353 & 0.115 & 2.253 & [1.729, 2.936] & & ~~27.31\\
6C**0956+4735 &  17.192 & -0.021 & 0.130 & 0.952 & [0.706, 1.284] & 1.026 & ~~27.39 & 27.47\\
6C**0957+3955 &  18.264 &  0.204 & 0.127 & 1.599 & [1.193, 2.143] & & ~~26.92\\
\\ 					   
6C**1003+4827 &  16.950 & -0.074 & 0.129 & 0.842 & [0.626, 1.133] & & ~~27.30\\
6C**1004+4531 &  17.183 & -0.023 & 0.130 & 0.948 & [0.703, 1.278] & & ~~26.41\\
6C**1006+4135 &  19.612 &  0.430 & 0.104 & 2.691 & [2.115, 3.423] & & ~~27.40\\
6C**1009+4327 &  20.101 &  0.490 & 0.095 & 3.089 & [2.479, 3.850] & 1.956 & ~~28.42 & 27.91\\
6C**1015+5334 &  18.516 &  0.253 & 0.124 & 1.790 & [1.344, 2.383] & & ~~27.42\\
\\ 			      	     				   
6C**1017+3436 &  18.972 &  0.334 & 0.117 & 2.157 & [1.647, 2.824] & & ~~27.53\\
6C**1018+4000 &  18.434 &  0.237 & 0.125 & 1.725 & [1.292, 2.303] & & ~~26.94\\
6C**1035+4245 &  17.250 & -0.0089& 0.130 & 0.980 & [0.727, 1.320] & & ~~26.96\\
6C**1036+4721 &  16.967 & -0.071 & 0.129 & 0.849 & [0.631, 1.143] & 1.758 & ~~27.02 & 27.81\\
6C**1043+3714 &  17.167 & -0.026 & 0.130 & 0.941 & [0.698, 1.268] & 0.789 & ~~26.98 & 26.79\\
\\ 			       	     				   
6C**1044+4938 &  18.472 &  0.244 & 0.125 & 1.754 & [1.317, 2.338] & & ~~27.48\\
6C**1045+4459 &  18.225 &  0.196 & 0.127 & 1.569 & [1.170, 2.105] & 2.571 & ~~27.12 & 27.66\\
6C**1048+4434 &  19.415 &  0.233 & 0.125 & 1.711 & [1.282, 2.284] & & ~~27.38\\
6C**1050+5440 &  19.715 &  0.443 & 0.103 & 2.773 & [2.188, 3.516] & & ~~27.79\\
\hline
\end{tabular}
\end{center}
\end{table*}

\begin{table*}
\begin{center}
\begin{tabular}{lllllllll}
\hline
\mc{1}{c}{Source} & \mc{1}{c}{$K$} & \mc{1}{c}{$\mu$} &
\mc{1}{c}{$\sigma$} & \mc{1}{c}{$z_{\rm{est}}$} & \mc{1}{c}{68\% C.I.}
& \mc{1}{c}{$z_{\rm{spectr}}$} & \mc{2}{c}{$\log_{10}L_{151}$}\\
\mc{1}{c}{} & \mc{1}{c}{} & \mc{1}{c}{} & \mc{1}{c}{} & \mc{1}{c}{} &
\mc{1}{c}{} & \mc{1}{c}{} & \mc{1}{c}{{\tiny est.}} &
\mc{1}{c}{{\tiny spec}.}\\
\hline 			       	     				   
6C**1052+4349 &  17.081 & -0.046 & 0.130 & 0.900 & [0.667, 1.213] & & ~~26.22\\
6C**1056+5730 &  17.295 &  0.00073 & 0.130 & 1.002 & [0.743, 1.351] && ~~27.09\\
6C**1100+4417 &  18.095 &  0.170 & 0.128 & 1.478 & [1.100, 1.987] & & ~~26.93\\
6C**1102+4329 &  19.661 &  0.436 & 0.104 & 2.732 & [2.150, 3.472] & 2.734 & ~~27.78 & 27.79\\
\\ 			       	     				   
6C**1103+5352 &  19.929 &  0.469 & 0.098 & 2.947 & [2.350, 3.695] & & ~~28.21\\
6C**1105+4454 &  17.729 &  0.092 & 0.129 & 1.237 & [0.918, 1.667] & & ~~26.77\\
6C**1106+5301 &  17.354 &  0.013 & 0.130 & 1.030 & [0.764, 1.390] & & ~~26.58\\
6C**1112+4133 &  18.044 &  0.159 & 0.129 & 1.443 & [1.073, 1.941] & & ~~26.87\\
6C**1125+5548 &  19.467 &  0.410 & 0.108 & 2.570 & [2.005, 3.295] & & ~~27.55\\
\\ 			       	     				   
6C**1132+3209 &  14.093 & -0.663 & 0.119 & 0.217 & [0.165, 0.286] & 0.231 & ~~24.84 & 24.90\\
6C**1135+5122 &  18.341 &  0.219 & 0.126 & 1.656 & [1.239, 2.215] & & ~~27.02\\
6C**1138+3309 &  18.014 &  0.153 & 0.129 & 1.423 & [1.058, 1.914] & & ~~27.05\\
6C**1138+3803 &  16.939 & -0.077 & 0.129 & 0.838 & [0.623, 1.127] & & ~~26.15\\
6C**1149+3509 &  18.729 &  0.292 & 0.121 & 1.958 & [1.482, 2.588] & & ~~27.15\\
\hline
\end{tabular}
\end{center}
{\caption{\label{tab:estimates_median} Redshift estimates for all the
    members of the 6C** sample. Column one lists the source
    names. Column two lists the \kband magnitudes measured in, or
    corrected to an 8-arcsec diameter aperture (see
    Section~\ref{sec:estimates}). Columns three and four list the
    probability density function (Eq.~\ref{eq:pdf_lognormal})
    parameters $\mu$ and $\sigma$, respectively. Column five lists the
    redshift estimates as obtained through Eq.~\ref{eq:mean}. Column
    six lists the 68\% confidence intervals as defined in
    Eq.~\ref{eq:ci}. Column seven lists the spectroscopic redshifts
    (references as in Table~\ref{tab:6cssummary1_median}). Columns
    eight and nine list the rest-frame 151 MHz radio luminosities
    (measured in units of W Hz$^{-1}$ sr$^{-1}$ and calculated
    assuming a power-law spectral index) based on the estimated and
    spectroscopic redshifts, respectively. Note: 6C**0737+5618 and
    6C**0935+4348 are not identified in $K$-band down to a 3$\sigma$
    limiting magnitude of $K \sim 21$~mag in a 8-arcsec diameter
    aperture (Paper I). Therefore, we do not extract probability
    density functions for these sources. We give lower limits for
    their redshift estimates, although we caution that they have been
    determined by extrapolation of our method to $K \,\,\simgt\,\,
    21$\,mag, where the $K-z$ diagram is not well-defined.}}
\end{table*}

\section{Comparison with spectroscopic redshifts} 
\label{sec:comparison}

\begin{figure} 
\centerline{
\psfig{figure=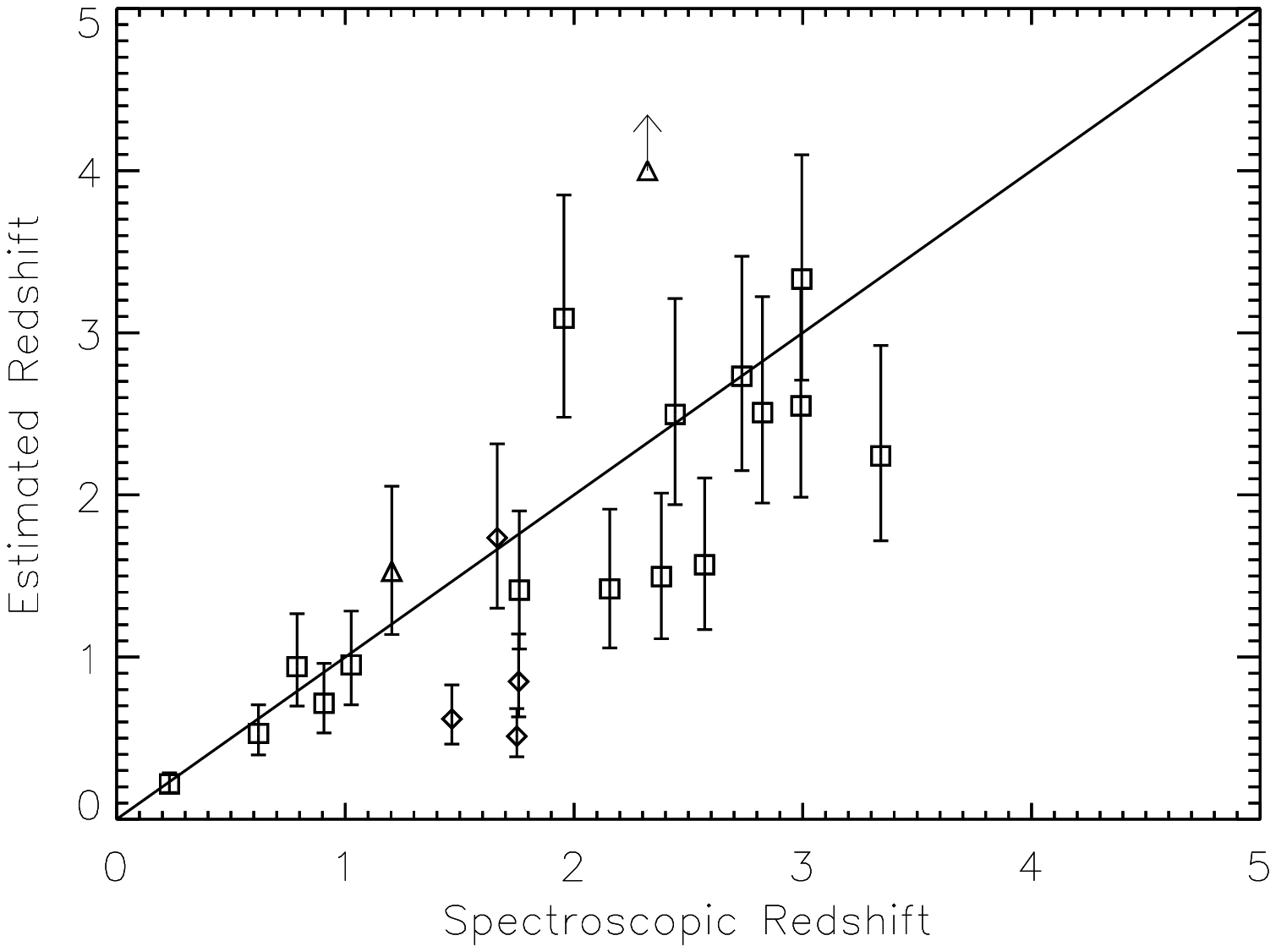,width=0.50\textwidth}} 
{\caption{\label{fig:comparison}  Comparison of estimated and
    spectroscopic redshifts for 6C** sources. The error bars show the
    asymmetric 68\% confidence interval about the best-fitting redshift
    estimate. The solid line is the $z_{\rm{est}}= z_{\rm{spec}}$
    line. Triangles represent the sources with uncertain redshifts,
    boxes the sources with secure redshifts, and diamonds the quasars.
    The triangle with an upward arrow represents the lower limit estimate
    for 6C**0935+4348 (see also the caption of Table~\ref{tab:estimates_median}). 
}}
\end{figure}

In Fig.~\ref{fig:comparison} we plot redshift estimates against
spectroscopic redshifts for the 6C** sample. It is apparent that the
majority of sources lie under the line of equality between estimated
and spectroscopic redshifts, implying a bias towards redshift
under-estimation. This is not unexpected given the level of
spectroscopic incompleteness of the sample. The subset of sources for
which we have spectroscopic redshifts is biased towards sources which
have strong emission lines and for which spectroscopy is easier to
obtain. Depending on their redshift, these are more likely to be the
sources in which emission line contamination to the \kband light is
more significant. Indeed, if we exclude the quasars, it can be seen
that the sources which show larger deviations from the equality line
are in excess at $z > 2$. As we have discussed in
Section~\ref{sec:emissionline}, this is consistent with an
unaccounted-for emission line contribution which systematically biases
some of the sources towards lower redshifts.  However, as it is clear
from Fig.~\ref{fig:pdfs}, the deviations are in all of the cases --
6C**0746+5445 ($z_{{\rm spec}} = 2.156$, Fig.~\ref{fig:pdfs}d);
6C**0832+5443 ($z_{{\rm spec}} = 3.341$, Fig.~\ref{fig:pdfs}h);
6C**0854+3500 ($z_{{\rm spec}} = 2.382$, Fig.~\ref{fig:pdfs}j) and
6C**1045+4459 ($z_{{\rm spec}} = 2.571$, Fig.~\ref{fig:pdfs}s) --
within the 95\% confidence interval upper-bound.

Not surprisingly, the quasars as a whole show the largest deviations
from the \mbox{$z_{\rm{est}}= z_{\rm{spec}}$} line. The $K-z$ diagram
does not hold for these objects and therefore their \kband magnitudes
make poor redshift estimators. Quasars can have a strong contribution
to their \kband light by the non-stellar continuum produced by the
AGN. For this reason, any attempted estimates via the $K-z$ diagram
will be invariably much lower than the true redshift. This is
particularly clear in Fig.~\ref{fig:pdfs}~(a,m,q), where it can be
seen that the estimates deviate by more than 2$\sigma$, and in one
case (6C**0922+4216) more than 3$\sigma$, from the true redshift. We
note, however, that the redshift of one of the quasars (6C**0928+4203
with $z_{{\rm spec}} = 1.664$, Fig.~\ref{fig:pdfs}n) is accurately
estimated with our method. Possibly because its broad Mg~II emission
is the result of scattered light from the broad line region and the
nucleus is heavily obscured (Paper I).

\begin{figure}
\centerline{
\psfig{figure=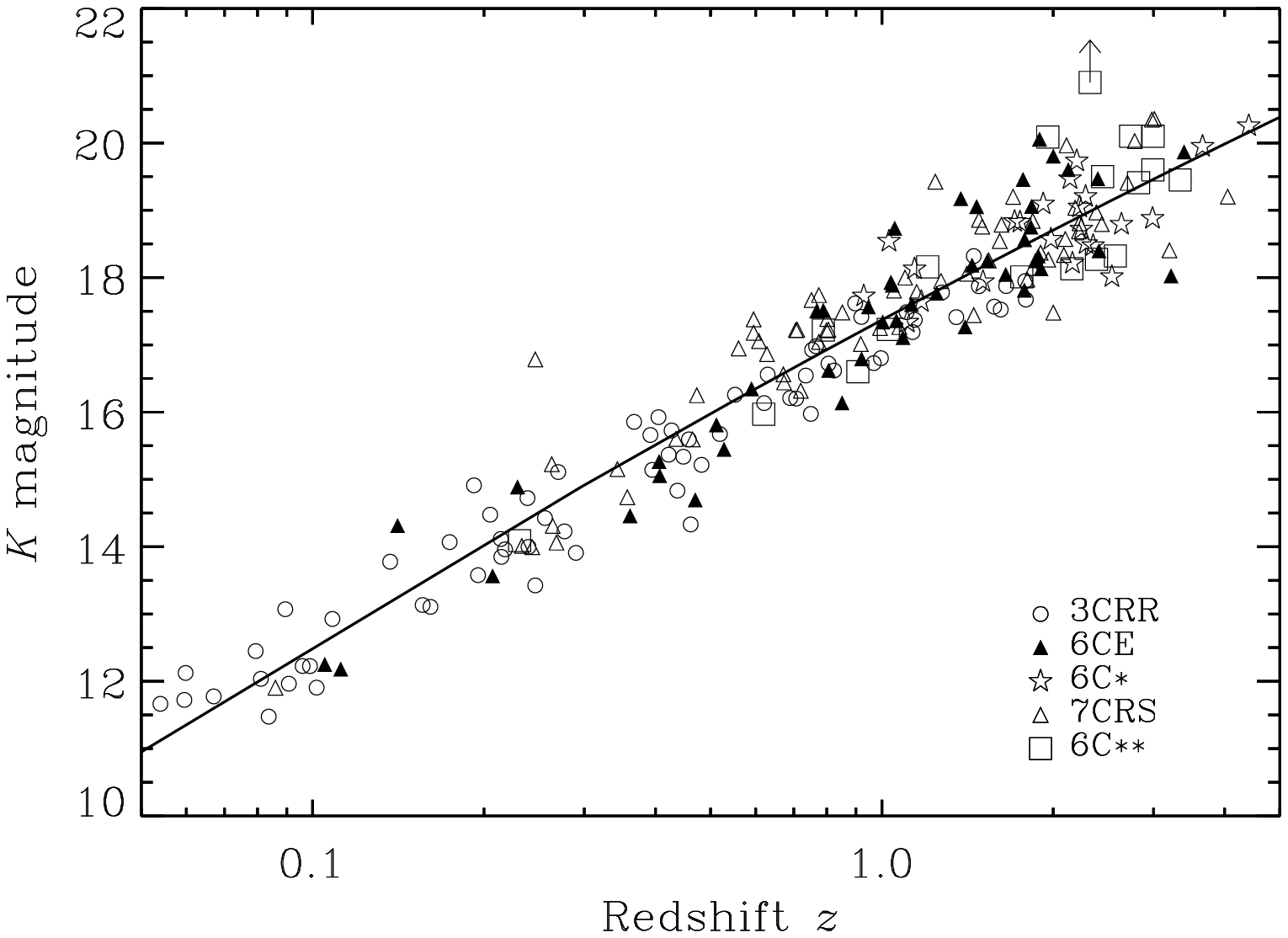,width=0.50\textwidth}} 
%\begin{center} 
%{\hbox to 0.4\textwidth{ \psfig{figure=fig6.eps,width=0.45\textwidth}}}
\caption{$K-z$ diagram of radio galaxies. The new spectroscopically
   confirmed 6C** data are plotted along with the 3CRR, 6CE, 6C* and
   7CRS. The solid line is the $K-z$ relation of Willott et
   al. (2003).  6C**0935+4348, the major 6C** $K-z $outlier, is
   represented by a box with an upward arraw, since this is also a
   $K$-magnitude lower limit.}
 \label{fig:kz_6cds}
%\end{center}
\end{figure}

 Two sources have photometric redshifts which are significantly
 overestimated by our method. One is 6C**1009+4327 ($z_{{\rm spec}} =
 1.956$, $z_{{\rm est}} = 3.089$), which is a very faint source ($K =
 20.5$\, mag in a 3-arcsec aperture) with spectrum showing weak
 Lyman-$\alpha$ emission. The other is 6C**0935+4348 ($z_{{\rm spec}}
 = 2.321$, $z_{{\rm est}} > 4.0$), which is one of the faintest
 sources in the sample, with $K > 21.7$~mag in a 3-arcsec
 aperture. This source is two magnitudes fainter than the mean $K-z$
 relation, which makes it a substantial $K-z$ outlier (see
 Fig.~\ref{fig:kz_6cds}). However, one caveat should be added here:
 the identification and redshift for this source are uncertain (Paper
 I).  Sources like this are very unusual, but do exist.  One of the
 7CRS galaxies, 5C7.178 at $z = 0.246$, is also $\sim 2$ magnitudes
 fainter than the mean $K-z$ relation (Willott et al. 2002b, 2003).

For both 6C**0935+4348 and 6C**1009+4327, the discrepancy between
 \kband magnitude and redshift suggests that the stellar content is
 not consistent with that of a massive host-galaxy. One possibility is
 that these sources are not yet fully formed (e.g. Jarvis, van
 Breukelen \& Wilman 2005). If, as a consequence of the
 `youth-redshift' degeneracy described by Blundell \& Rawlings (1999),
 we are preferentially observing young sources ($\ltsim 10^{7}$~yr),
 then in some cases this could be the first instance of accretion
 activity in those sources. As such, this may trigger, or occur
 simultaneously with the initial star formation process in the host
 galaxy (e.g. Willott, Rawlings \& Jarvis 2000; Willott et
al. 2002a). If this is the case, it is perhaps not too surprising that
 their redshifts are overestimated by our method. As discussed by
 Willott et al. (2003), the best-fitting $K-z$ relation is close to
 the expected $K$-magnitude evolution of a galaxy of local luminosity
 3~$L_{*}$ which forms all of its stars in an instantaneous burst at
 $z_{f} = 10$.

However, 5C7.178, the 7CRS prominent outlier to the $K-z$ relation, is
at low redshift and therefore unlikely to be a forming galaxy. This
suggests that the 'young galaxy' hypothesis is not the only cause of
underluminous outliers. The existence of a clear bright envelope to the
K-z relation is presumably intimately related to the exponential
cut-off in the galaxy Schechter function, whereas a tail to fainter
magnitudes would be expected in any model in which powerful jets can,
if only rarely, be associated with underluminous galaxies.

Apart from the objects so far discussed, the majority of sources have
been assigned with a reasonably well-constrained redshift estimate, in
that the true redshift lies within the 68 per cent confidence interval
about the best-fitting redshift estimate. Our method seems therefore
to be fairly robust whenever emission-line and/or non-stellar
contributions to the \kband light can be neglected. However, we note
that this could be a selection effect resulting from the spectroscopic
incompleteness of the sample. Among the subsample of objects for which
we do not have spectroscopic redshifts there could possibly be other
sources like 6C**1009+4327 and 6C**0935+4348, that have redshift
estimates significantly higher than their true redshifts.

We now compare the near-infrared Hubble diagram of the 6C** radio
galaxies (for which we have a spectroscopic redshift) with the 3CRR,
6CE, 6C* and 7CRS samples (see Fig.~\ref{fig:kz_6cds}).  We find that
the 6C** objects follow the same $K-z$ relation as the combined
3CRR/6CE/6C*/7CRS data, and have a rms dispersion of $\simeq 0.59$~mag
over all redshifts if 6C**0935+4348 is excluded. This is consistent
with the dispersion value used in our modelling
(Section~\ref{sec:kz_model}). If 6C**0935+4348, the prominent $K-z$
outlier, is included in this calculation, then the dispersion becomes
$\simeq 0.75$~mag.

\begin{figure}
\begin{center} 
{\hbox to 0.4\textwidth{ \epsfxsize=0.45\textwidth
\epsfbox{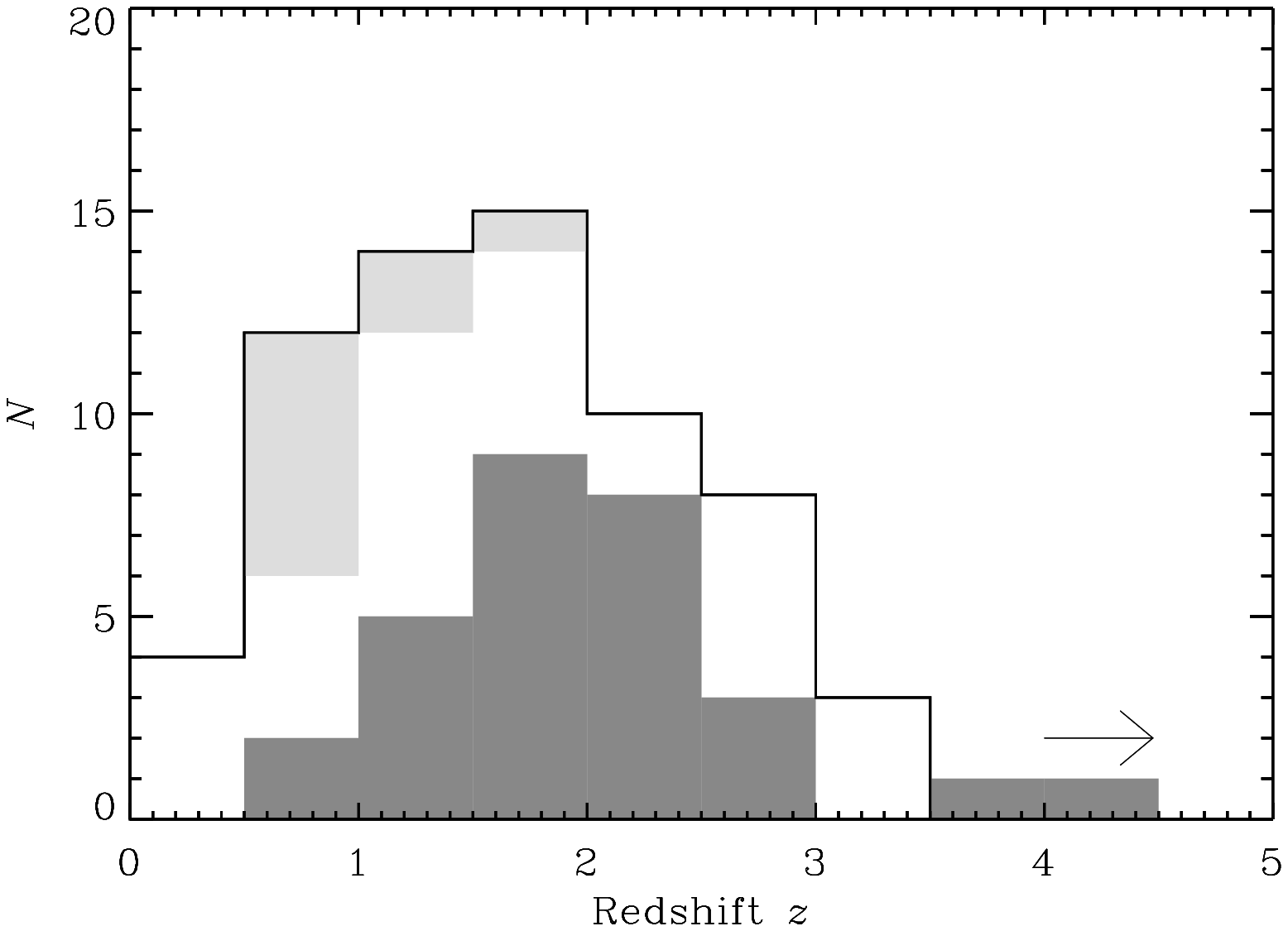}}}
\caption{Estimated redshift distribution of the sources in the 6C**
  sample (solid line). The bin width is $\Delta z = 0.5$. The dark
  shadowed region is the distribution of sources in the 6C* sample
  (Jarvis et al. 2001b). The arrow above the $4.0 < z <
  4.5$ bin represents the lower limit redshift estimates for
  6C**0737+5618 and 6C**0935+4348. These are the two 6C** sources which are not detected 
  in $K$-band (see also the
  caption of Table~\ref{tab:estimates_median}). The light shadowed
  region shows the locus of the nine quasars of which we know about in the
  6C** sample (listed in table~9 of Paper I).}
 \label{fig:histzest}
\end{center}
\end{figure}

\section{The estimated redshift distribution}
\label{sec:est-zdist}

The estimated redshift distribution for the entire 6C** sample is
presented in Fig~\ref{fig:histzest}. The median estimated redshift is
$z \simeq 1.6$. It is informative to compare these results with the
redshift distribution of the 6C* sample (Jarvis et al. 2001b) which,
as mentioned before, has been filtered in a very similar way. The 6C*
sample has a median redshift of $z \simeq 1.9$ and a redshift
distribution skewed towards $z > 2$. This is a direct result of the
filtering criteria applied to it.

Although the distribution of 6C** sources does have a tail to high
redshift, the great majority of sources are at \mbox{$z <2$}, in clear
contrast to what should be expected. However, and in light of
Section~\ref{sec:comparison}, we have to consider what is likely to be
the major source of bias in our redshift estimation: the presence of
quasars in the sample. These sources will be the ones which will be
skewing the distribution the most towards lower redshifts.  The
redshift estimates of the three quasars with unresolved
identifications (6C**0714+4616, 6C**0922+4216 and 6C**1036+4721), for
which we have a spectroscopic redshift, are under-estimated by an
average factor of 0.4.  The major cause for concern is that there are
quasars for which we do not have spectroscopic information.  On the
basis of \kband imaging alone, we assume that the further following
sources are quasars (given that they have bright unresolved
identifications; Paper I): 6C**0849+4658, 6C**1003+4827,
6C**1052+4349, 6C**1056+5730 and 6C**1138+3803. This leaves us with
nine quasars\footnote{These include 6C**0928+4203, which does not
appear unresolved in our near-infrared imaging, but shows broad
emission lines in its spectrum (Paper I). We recall that this source
has a redshift estimate which is in good agreement with its
spectroscopic redshift (Section~\ref{sec:comparison}).}, eight of
which have redshift estimates in the range \mbox{$0.5 < z
\,\,\ltsim\,\, 1.0$}.  These alone represent a significant source of
the bias towards low-redshift which we witness in
Fig~\ref{fig:histzest}.

Removing the quasars from the distribution results in a median
estimated redshift of $z \simeq 1.7$.  This is similar to what is
found for 6C* and much higher than the median redshift of similar,
unfiltered samples (e.g. the 7CRS sample, at the same flux-density
limit, with $z \approx 1.1$). This result confirms the efficiency of
the filtering criteria, used for 6C**, in excluding low-redshift
sources.  Furthermore, we recall that some sources may also have their
redshifts systematically under-estimated, due to emission-line
contamination to their \kband magnitudes
(Sections~\ref{sec:emissionline} and \ref{sec:comparison}). Thus, it
is quite plausible that the real redshift distribution has a slightly
higher median redshift.

The fraction of quasars in the 6C** sample (9 quasars out of 68
objects) is higher than that in the 6C* sample. There are only two
quasars out of the 29 6C* objects (Jarvis et al. 2001b). Although the
difference is barely significant given the small numbers involved, it
is possibly influenced by small differences in the selection criteria,
such as the tighter size constraint or, more importantly, the lower
frequency spectral index cut. For 6C*, the steep spectral index
constrain is evaluated between 151\,MHz and 4.85\,GHz, thus excluding
objects with prominent flat-spectrum cores, i.e. quasars.  Because the
6C** spectral index cut goes up to only 1.4\,GHz, this effect is less
pronounced.

\section{The RLF model of steep-spectrum radio sources}
\label{sec:JarvisRLF}

Jarvis et al. (2001c) investigated the radio luminosity function for
the most radio luminous low-frequency selected sources. The data used
in their analysis were drawn from the complete 3CRR, 6CE and filtered
6C* samples by imposing a lower limit in radio luminosity.  Only the
sources which lie in the top-decade in $\log_{10}\, L_{151}$ were
considered\footnote{This meant: $\log_{10}\, L_{151} \geq 27.63$ in
their cosmology I ($\Omega_{M} = 1.0$, $\Omega_{\Lambda} = 0.0$,
$H_{0} = 50$ km~s$^{-1}$~Mpc$^{-1}$) and $\log_{10}\, L_{151} \geq
27.90$ in their cosmology II ($\Omega_{M} = 0.3$, $\Omega_{\Lambda} =
0.7$, $H_{0} = 50$ km~s$^{-1}$~Mpc$^{-1}$), the two sets of
cosmological parameters considered in their analysis.}. Focusing on
only the most luminous sources provided the largest possible baseline
in redshift for the samples considered, while minimizing the role
played (in the modelling procedure) by the intrinsic correlations
between sample parameters, such as luminosity - spectral index and
linear size - spectral index correlations, which are inherent to radio
samples (Blundell, Rawlings \& Willott 1999).

The filtered 6C* sample represented the greatest advance of this study
over earlier work on the high-redshift RLF of low-frequency selected
radio sources (Willott et al. 2001).  However, its use required that
the effects of the selection criteria, namely small angular size and
steep radio spectral index, and in particular the fraction of sources
which were missing from the survey, were taken into consideration in
the modelling of the RLF. This led to a parameterisation which was
separable in luminosity and redshift (as in Jarvis \& Rawlings 2000
and in Willott et al. 2001), and also incorporated distributions in
radio spectral shape and linear size, i.e.
\begin{eqnarray}\label{eqn:rlf}
& \rho (L_{151},z,a_{1},a_{2},D) = & \rho_{\circ} \times \rho_{L}
(L_{151}) \times \rho (z)  \nonumber \\ & & \times \rho_{a}
(a_{1},a_{2}) \times \rho_{D} (D),
\end{eqnarray}
where $\rho_{\circ}$ is the normalising factor and a free parameter
measured in units of Mpc$^{-3}$, $\rho_{L}(L_{151})$, $\rho (z)$,
$\rho_{a} (a_{1},\,a_{2})$ and $\rho_{D} (D)$ are dimensionless
distribution functions in radio luminosity, redshift, spectral shape
parameters and projected linear size, respectively.

The full form of these distributions and their modelling are described
in detail in Jarvis et al. (2001c).  We now comment on the shape of
the redshift distribution, for which Jarvis et al. (2001c) tested
three different models.  The model favoured by their maximum
likelihood analysis (model C) uses a 1-tailed Gaussian to parameterise
the low-redshift co-moving space density and a power-law distribution
at high redshift, i.e
\begin{equation}\label{eqn:rhozC}
\rho_{C}(z) = \left\{ \begin{array} {l@{\quad:\quad}l} \exp \left
\{-\frac{1}{2} \left( \frac{z-z_{\circ}}{z_{1}} \right)^{2} \right \}
& z \leq z_{\circ} \\ \left (\frac{1
+ z}{1 + z_{\circ}}\right )^{\eta} &  z > z_{\circ} ,  \end{array}
\right.
\end{equation}
where $z_{\circ}$ is the `break' redshift where the model switches
from the low- to the high-redshift form, $z_{1}$ is the characteristic
width of the half-Gaussian and $\eta$ is the power-law exponent
describing the high-redshift co-moving space density. The other models
tested by Jarvis et al. (2001c) were: model A -- parameterised as a
single Gaussian distribution in redshift; and, model B -- a 1-tailed
Gaussian which becomes constant and equal to unity beyond the Gaussian
peak, i.e. the same as model C with $\eta$ fixed at zero. The
advantage of model C over the other models is that: (i) it breaks the
symmetry between low- and high-redshift evolution which is forced by
model A (and for which there is no physical basis), and (ii) allows
for freedom in the evolution at high-redshift which is not possible
with model B.  Model C, is therefore more useful in terms of assessing
the form of the evolution of the co-moving space density at high
redshift.

Jarvis et al. (2001c) found a best-fitting power-law exponent $\eta =
-0.06$ (in their Cosmology II), implying a constant co-moving space
density beyond a peak redshift of $z_{\circ} = 2.15$ to an
indeterminable redshift.  A steep decline has been ruled out by their
analysis at the $\sim 4 \sigma$ level, but the form of the evolution
at high redshift remained unresolved with an uncertainty encompassing
both moderate declines and continuing shallow inclines.

\section{Comparison with the RLF of steep-spectrum radio sources}\label{sec:RLF}

To investigate the co-moving space density of high-redshift
steep-spectrum radio sources, we now compare the redshift estimates
derived from the 6C** data with the RLF of low-frequency selected
radio sources of Jarvis et al. (2001c).

\subsection{Selecting the most luminous 6C** sources}
\label{sec:most-luminous-sources}

\begin{figure}
  \begin{center}
    \begin{minipage}[c]{1.0\linewidth}
      \epsfxsize=3.3in\epsfbox{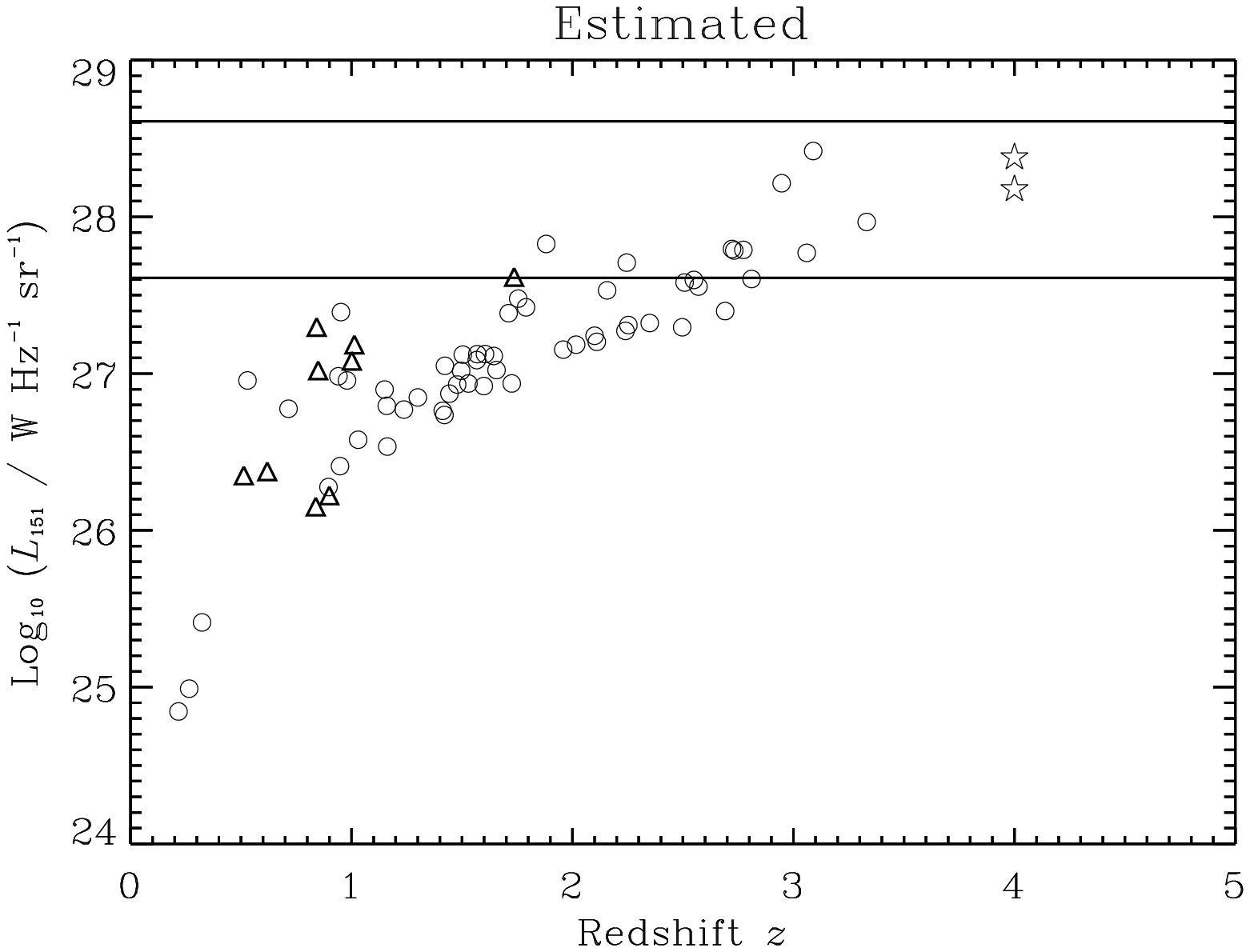}   
    \end{minipage}
    \begin{minipage}[c]{1.0\linewidth}
      \epsfxsize=3.3in\epsfbox{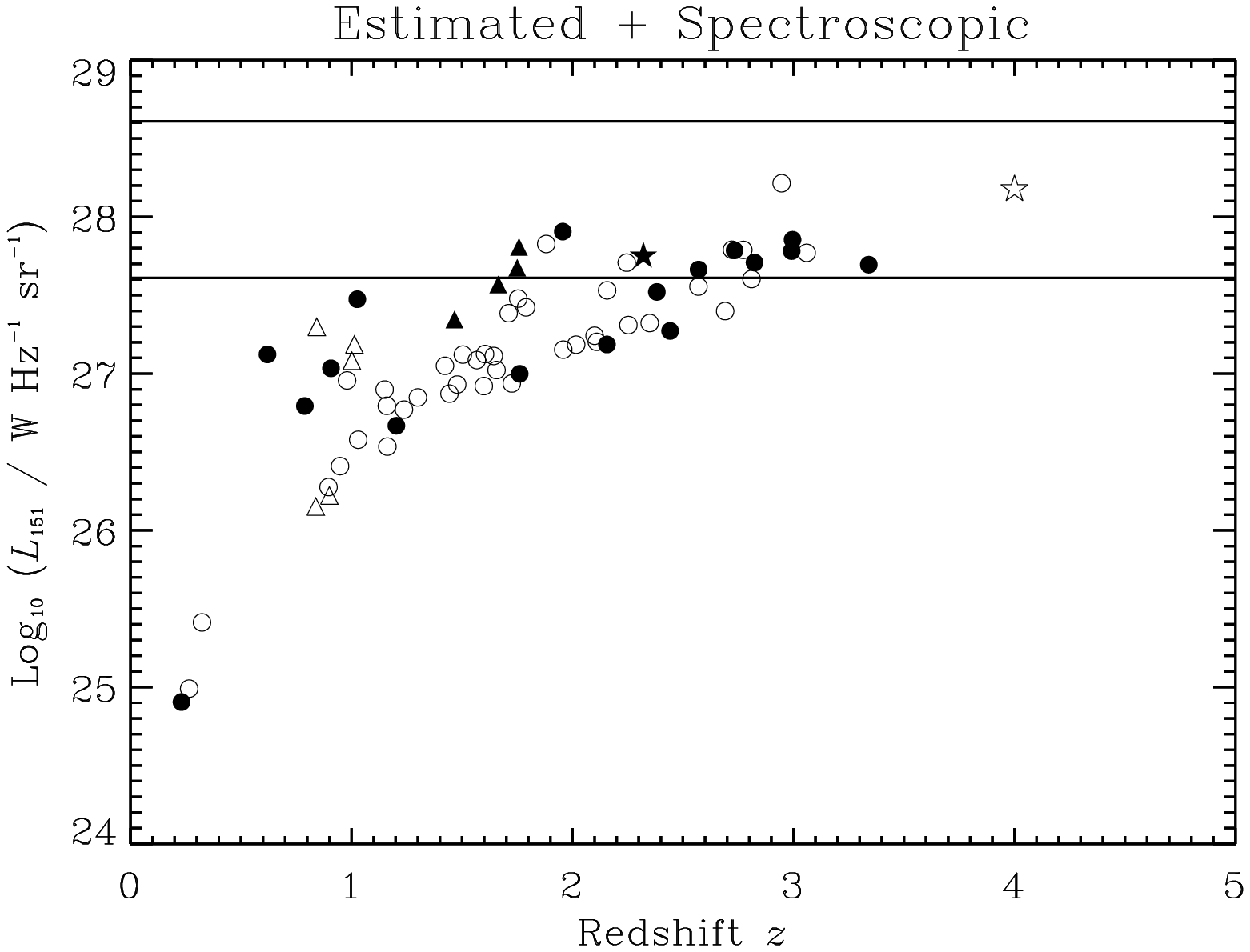}   
    \end{minipage}
\end{center}
\caption{The estimated 151-MHz luminosity-redshift plane for the 6C**
  sample. The radio galaxies are represented by circles and the
  quasars by triangles. In the bottom-panel spectroscopic redshifts
  are taken into consideration (filled symbols).  In both plots the
  star symbols represent the sources which were not detected in \kband
  (down to \mbox{$K \simeq 21$~mag} in an 8-arcsec aperture), i.e.
  6C**0737+5618 and 6C**0935+4348.  The area between the horizontal
  lines is the region which contains the `most luminous' sources
  according to our definition, and which is equivalent to the
  top-decade in luminosity defined by Jarvis et al. (2001c).  These
  plots are for $\Omega_{M} = 0.3$, $\Omega_{\Lambda} = 0.7$, $H_{0} =
  70$ km~s$^{-1}$~Mpc$^{-1}$.}
 \label{fig:pzestimated}
\end{figure}

To be able to compare the 6C** data with the RLF model of Jarvis et
al. (2001c), we must select the most radio luminous sources in the
sample in a fashion that is equivalent to that of Jarvis et
al. (2001c).  For that we use the definition of top-decade in
cosmology II of Jarvis et al. (2001c) and translate its lower radio
luminosity limit to $\log_{10}\, L_{151} \geq 27.61$ in the cosmology
used in this paper ($\Omega_{M} = 0.3$, $\Omega_{\Lambda} = 0.7$,
$H_{0} = 70$ km~s$^{-1}$~Mpc$^{-1}$). This means that, based strictly
on the estimated $L_{151}-z$ diagram (Fig~\ref{fig:pzestimated}:
top-panel), 12 of the 6C** sources are selected. However, if we
consider also the spectroscopic redshifts available, i.e. by replacing
estimated redshifts with spectroscopic ones where these are available
(Fig~\ref{fig:pzestimated}: bottom-panel), then there are 17 sources
in the top decade of radio luminosity. These sources are listed in
Table~\ref{tab:top-decade}, and include 6C**0737+5618. The inclusion
of this source is discussed in more detail in
Section~\ref{sec:pdfdistribution}. We note that two of the quasars
with under-estimated redshifts (and luminosities) are now in the
top-decade. A few of the quasars without spectra are also expected to
move into the top-decade once their redshifts are known.

\begin{table}
\begin{center}
\begin{tabular}{l|cc|cc|c}
\hline
\mc{1}{c|}{Source} & \mc{2}{|c|}{Estimated} & \mc{2}{|c|}{Spectroscopic} & \mc{1}{c}{Cl.}\\
\mc{1}{c|}{name} & \mc{1}{|c}{$z_{\rm est}$} & \mc{1}{c|}{$\scriptstyle\log_{10}L_{151}$} & \mc{1}{c}{$z_{\rm spec}$} & \mc{1}{c|}{$\scriptstyle\log_{10}L_{151}$} &\\
\hline
6C**0737+5618 & $\scriptstyle>$4.0 & $\scriptstyle>$28.18 &        &       &  \\
6C**0744+3702 &	2.549  &  -----      & 2.992  & 27.78 & G\\			      
6C**0754+5019 & 3.331  & 27.97    & 2.996  & 27.85 & G\\
6C**0801+4903 & 2.722  & 27.79    &        &       &  \\
6C**0824+5344 & 2.507  &  -----      & 2.824  & 27.71 & G\\   	       			      
6C**0832+5443 & 2.240  &  -----      & 3.341  & 27.69 & G\\	       			      
6C**0902+3827 & 2.246  & 27.71    &        &       &  \\
6C**0909+4317 & 1.881  & 27.83    &        &       &  \\
6C**0922+4216 & 0.512  &  -----	  & 1.750  & 27.67 & Q\\      			      
6C**0925+4155 & 3.060  & 27.77    &        &       &  \\
6C**0928+4203 & 1.736  & 27.62    & 1.664  &  -----   & Q\\
6C**0935+4348 & $\scriptstyle>$4.0 & $\scriptstyle>$28.38    & 2.321  & 27.75 & G\\
6C**1009+4327 & 3.089  & 28.42    & 1.956  & 27.91 & G\\
6C**1036+4721 & 0.849  &  -----      & 1.758  & 27.81 & Q\\	       			      
6C**1045+4459 & 1.569  &  -----      & 2.571  & 27.66 & G\\	       			      
6C**1050+5440 & 2.773  & 27.79    &        &       &  \\
6C**1102+4329 & 2.732  & 27.78    & 2.734  & 27.79 & G\\
6C**1103+5352 & 2.947  & 28.21    &        &       &  \\
\hline
\end{tabular}
\end{center}
{\caption{\label{tab:top-decade} Sources selected from the 6C** sample
    which correspond to our top-decade definition (see
    Section~\ref{sec:most-luminous-sources}). The rest-frame 151~MHz
    luminosities are in units of W~Hz$^{-1}$~sr$^{-1}$ and were
    calculated assuming a simple power-law spectral index. The third
    column corresponds to a top-decade based strictly on the estimated
    redshift distribution. In this column the solid horizontal lines
    are used for sources which, based on their redshift estimates,
    fall below the lower luminosity limit. However, these sources are
    included in the top-decade if spectroscopic information (in
    columns 4--5) is taken into account. One source -- 6C**0928+4203,
    is excluded. The classifications (\mbox{G -- radio galaxy};
    \mbox{Q -- quasar}) are given only for sources which have optical
    spectroscopy (Paper I and references therein) and are based on
    these data.}}
\end{table}

\subsection{Analysis and discussion}

We consider two ways of constructing the binned redshift distribution
of 6C** sources in the top-decade of luminosity. In
Section~\ref{sec:best-fitting}, we use spectroscopic redshifts when
available, and otherwise the best-fitting redshift estimates, i.e. we
consider the sources listed in Table~\ref{tab:top-decade} (see also
Fig~\ref{fig:pzestimated}: bottom-panel).  In
Section~\ref{sec:pdfdistribution}, we take into account the redshift
probability distribution of each source. Both redshift distributions
are then compared with the expected redshift distribution given Jarvis
et al. (2001c) model RLF.

\subsubsection{The predicted redshift distribution}
\label{sec:predicted}

The predicted redshift distribution is obtained by integrating the
model C RLF of Jarvis et al. (2001c), normalized to the sky area and
evaluated for the radio selection criteria of the 6C** sample, i.e.
\begin{eqnarray}\label{eqn:dN/dz}
  \frac{{\rm d}N}{{\rm d}z} (z) &  = &  \int\!\!\int\!\!\int\!\!\int\!
\rho(L_{151},\,z,\,a_{1},\,a_{2},\,D) \nonumber \\ & & \times  \Omega(L_{151},\,z,\,a_{1},\,a_{2},\,D) \nonumber \\ 
 & & \times \frac{{\rm d}V}{{\rm d}z} \,{\rm 
d}(\log_{10}L_{151})\,{\rm d}a_{1}\, {\rm d}a_{2}\, {\rm d}(\log_{10}D)
\end{eqnarray}
where $\rho(L_{151},\,z,\,a_{1},\,a_{2},\,D)$ is the complete radio
luminosity function Jarvis et al. (2001c);
$\Omega(L_{151},\,z,\,a_{1},\,a_{2},\,D)$ is the sky area available
for the 6C** sample; and, $({\rm d}V/{\rm d}z)$ is the differential
co-moving volume element.  The factor
$\Omega(L_{151},\,z,\,a_{1},\,a_{2},\,D)$ assumes a value of 0.421~sr
(the sky area of 6C**) or zero, depending on whether or not a source
with a given a set of the parameters $L_{151},\,z,\,a_{1},\,a_{2}$ and
$D$ meets the sample selecting criteria.  The lower and upper limits
of the integral are $27.61 \leq \log_{10}L_{151} \leq 28.61$, $-2.2
\leq a_{1} \leq 1.0$, $-0.4 \leq a_{2} \leq 0.2$ and $-0.3 \leq
\log_{10} D \leq 4.0$. The integral was evaluated numerically, between
$0 \leq z \leq 8$, using a Monte Carlo method, over $10^{7}$ random
points uniformly distributed in the 4-dimensional parameter space.

Because we use a different set of cosmological parameters than that of
Jarvis et al. (2001c), we re-evaluated the luminosity function by
using the relation from Peacock (1985):
\begin{equation}
\rho_{1}(L_{1},z)\frac{{\rm d}V_{1}}{{\rm d}z} = \rho_{1}(L_{2},z)\frac{{\rm d}V_{2}}{{\rm d}z}
\label{eq:rlf}
\end{equation}
where $\rho_{1}$ and $\rho_{2}$ are the luminosity functions for two
different cosmologies, and, $L_{1}$ and $L_{2}$ are the luminosities
derived from the flux density, redshift and proper motion distance in
each of the two different cosmologies.  We used the RLF model
evaluated in Cosmology II of Jarvis et al. (2001c) to obtain the RLF
used in Eq.~\ref{eqn:dN/dz}.

The best-fitting parameter log$_{10} D_{o}$, the peak of the Gaussian
distribution in $\log_{10}$ of the projected linear size $D$ (eq.~7 of
Jarvis et al. 2001c), was also redefined due to the change in the set
of cosmological parameters used. Since this is just a change in the
Hubble constant, the relation between the linear sizes in the two
cosmologies is then simply proportional to the ratio of their
respective Hubble constants. For model C, log$_{10} D_{o}$ becomes
1.97 in the cosmology used in this paper. All the other parameters
remained the same (table~3 of Jarvis et al. 2001c).

\begin{figure}
  \begin{center}
    \begin{minipage}[c]{1.0\linewidth}
      \epsfxsize=3.3in\epsfbox{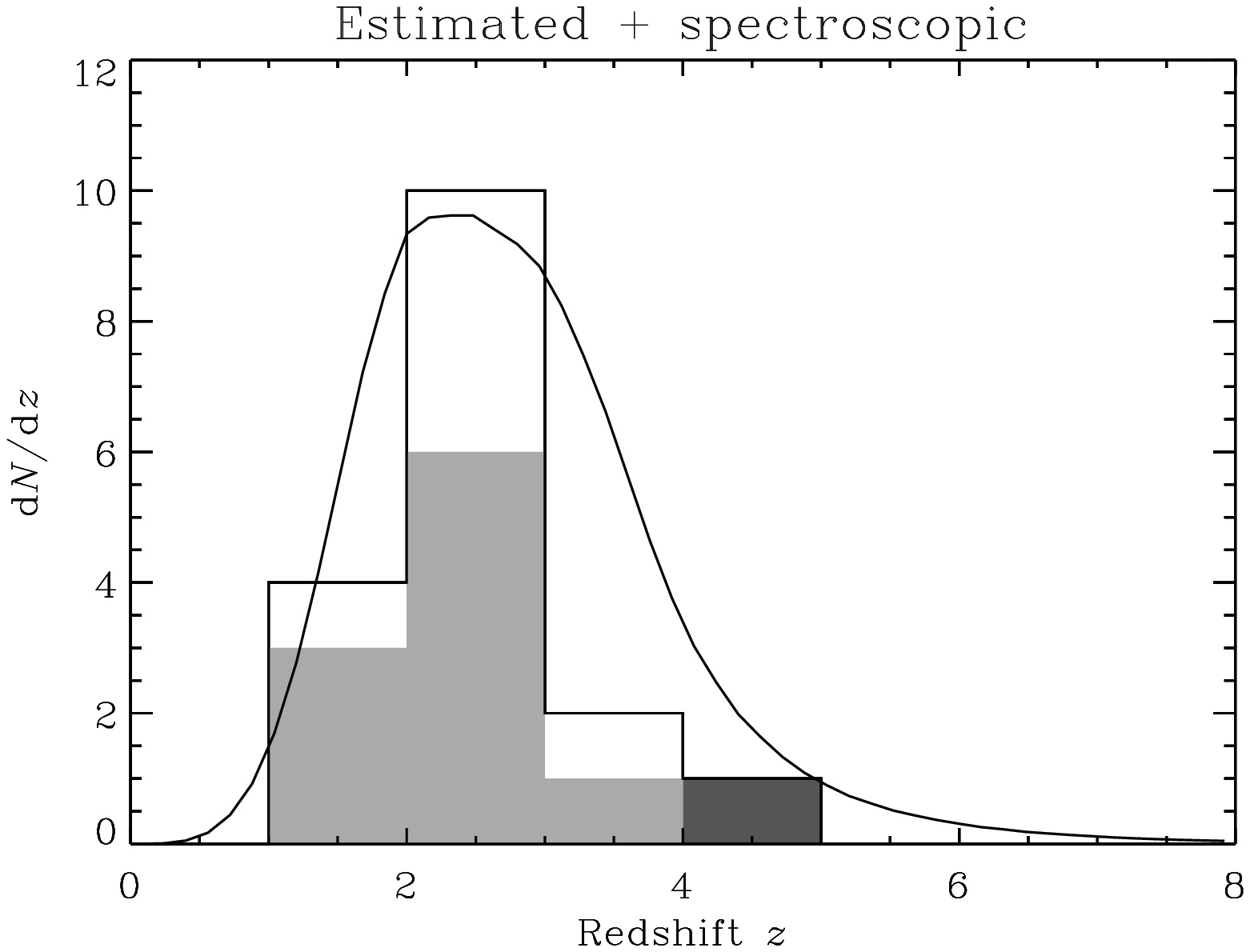}   
    \end{minipage}
  \end{center}
\caption{The histogram shows the number of sources in the 6C** sample
  with $\log_{10} L_{151} \geq 27.61$, drawn from the `estimated +
  spectroscopic' $L_{151}-z$ diagram (Fig.~\ref{fig:pzestimated}:
  bottom-panel), binned in redshift with bin width $\Delta z = 1$.
  The light shaded region represents the sources which have
  spectroscopic redshifts.  The solid line shows Jarvis et al. (2001c)
  model C predictions ($\eta = -0.06$, constant co-moving space
  density) for the redshift distribution
  (Section~\ref{sec:predicted}).  The dark shaded region represents
  6C**0737+5618, for which the redshift estimate is a lower limit
  based on extrapolation of the $K-z$ relation (see also the caption
  of Table~\ref{tab:estimates_median}).}
 \label{fig:redshift}
\end{figure}

\subsubsection{Using the best-fitting redshift estimates}
\label{sec:best-fitting}

The redshift distribution of 6C** sources in the top-decade of
luminosity, based on their best-fitting redshift estimates ($z_{\rm
est}$) and on the spectroscopic redshifts available, is shown in
Fig.~\ref{fig:redshift}. In general, the model redshift distribution is
a fairly good approximation to the data. There is excellent agreement
at $z < 2$, and it is also apparent that: (i) there is a slight
deficit of sources predicted at $ 2 < z < 3$; and, (ii) there is an
excess of predicted sources at $z > 3$.  However, we note that it is
at these redshifts that the effects of the small number statistics
become more important. Moreover, our redshift estimation method relies
on the limited statistics of the 3CRR, 6CE, 7CRS and 6C* radio
galaxies at $z > 3$, where the scatter in the $K-z$ diagram is still
poorly defined.

\begin{figure}
  \begin{center}
    \begin{minipage}[c]{1.0\linewidth}
      \epsfxsize=3.0in\epsfbox{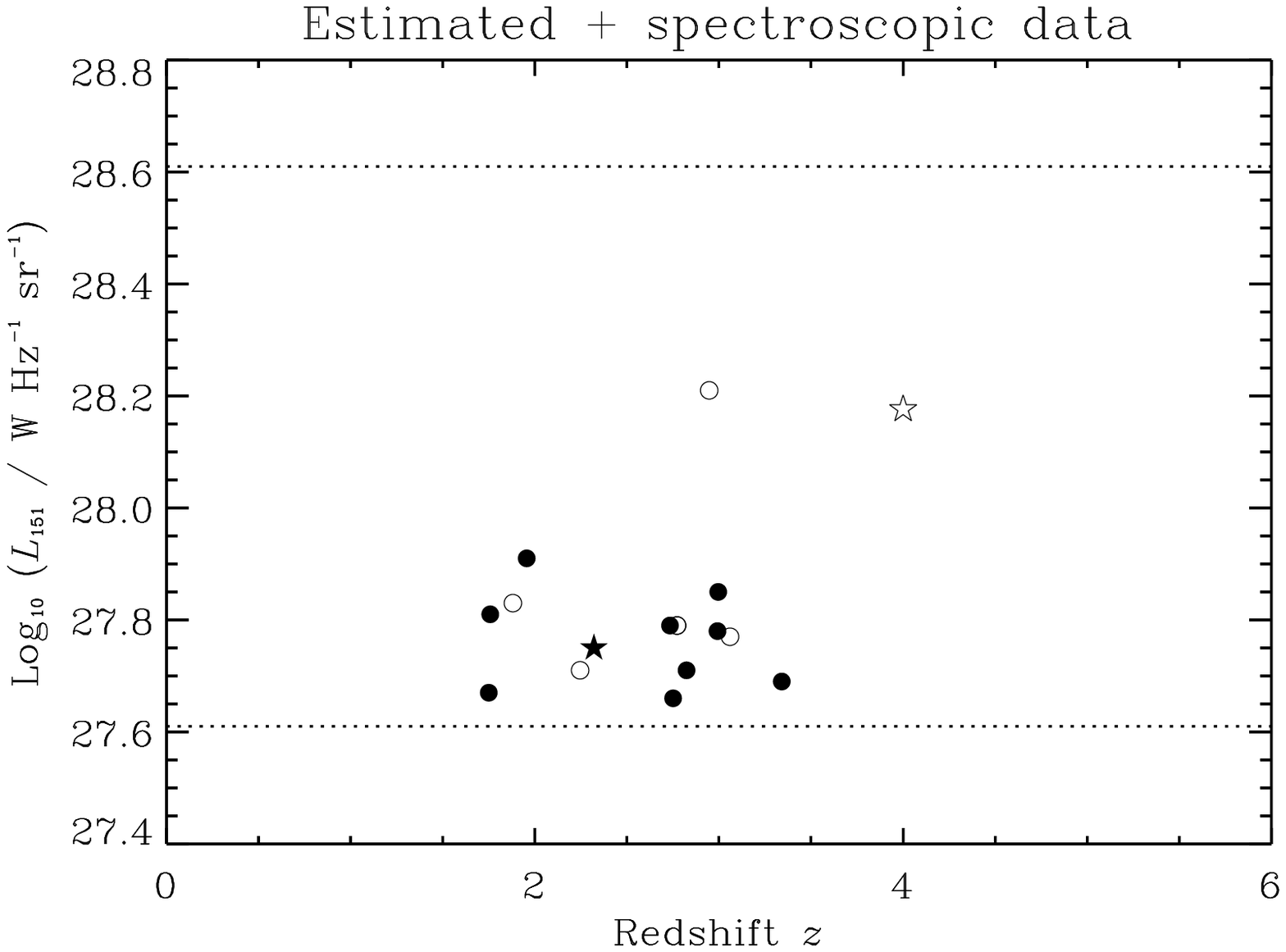}   
    \end{minipage}  
   \begin{minipage}[c]{1.0\linewidth}
      \epsfxsize=3.0in\epsfbox{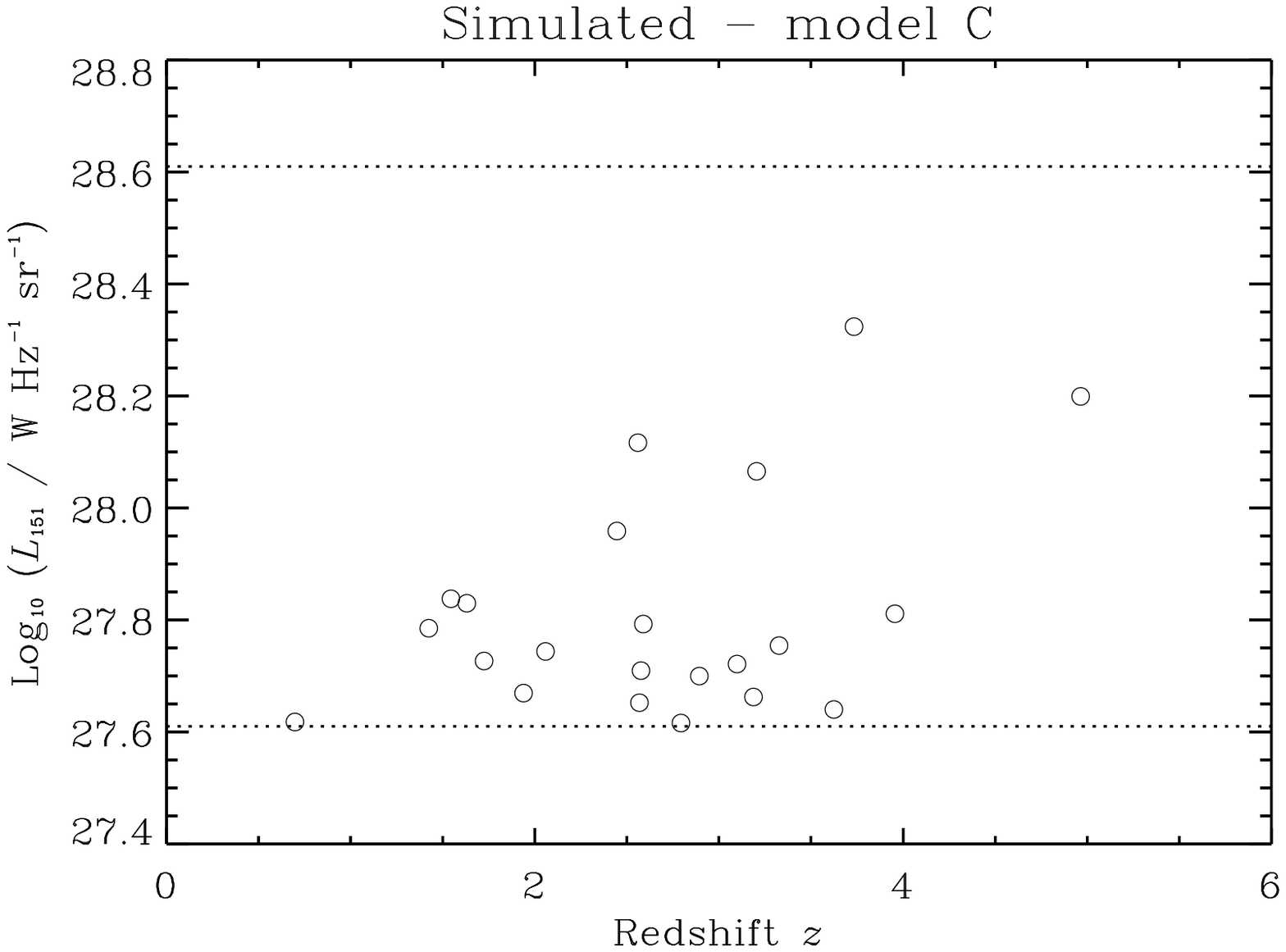}   
    \end{minipage}\
  \end{center}
\caption{The top panel shows the estimated top-decade (delimited by
  the dotted horizontal lines) of the $L_{151}-z$ plane for the 6C**
  sample. The spectroscopic redshifts are included (by replacing
  estimated redshifts with spectroscopic ones when these are
  available), and are represented by the filled symbols. The star
  symbol represents 6C**0737+5618. The bottom panel shows a simulation
  of the top-decade of the $L_{151}-z$ plane for a sample with the
  same flux limit, sky area and radio selection criteria as 6C**. This
  simulation was generated using the best-fitting model C of Jarvis et
  al. (2001c), translated to our adopted cosmology.}
\label{fig:pz:top-decade}
\end{figure}

Similar comments can be made when we use the RLF model to create an
artificial $L_{151}-z$ plane and compare it with the one estimated
from the data, including the spectroscopic redshifts
(Fig.~\ref{fig:pz:top-decade}). In general, it is apparent that the
model predicts more sources (22) than the ones that are present in the
estimated+spectroscopic diagram (17 sources). However, we caution
that, since both the estimated data and the simulations have
independent Poisson errors, this comparison can only be regarded in a
qualitative way.

To explore the significance of these results, we apply the
Kolmogorov-Smirnov test to compare the redshift distribution derived
from data and the predicted model distribution
(Fig.~\ref{fig:redshift}). We find that they are not significantly
different, with probability $p = 0.29$.  Thus, we can conclude that
the data are consistent with Model C of Jarvis et al. (2001c), the
constant co-moving space density model.

\begin{figure}
  \begin{center}
    \begin{minipage}[c]{1.0\linewidth}
      \epsfxsize=3.3in\epsfbox{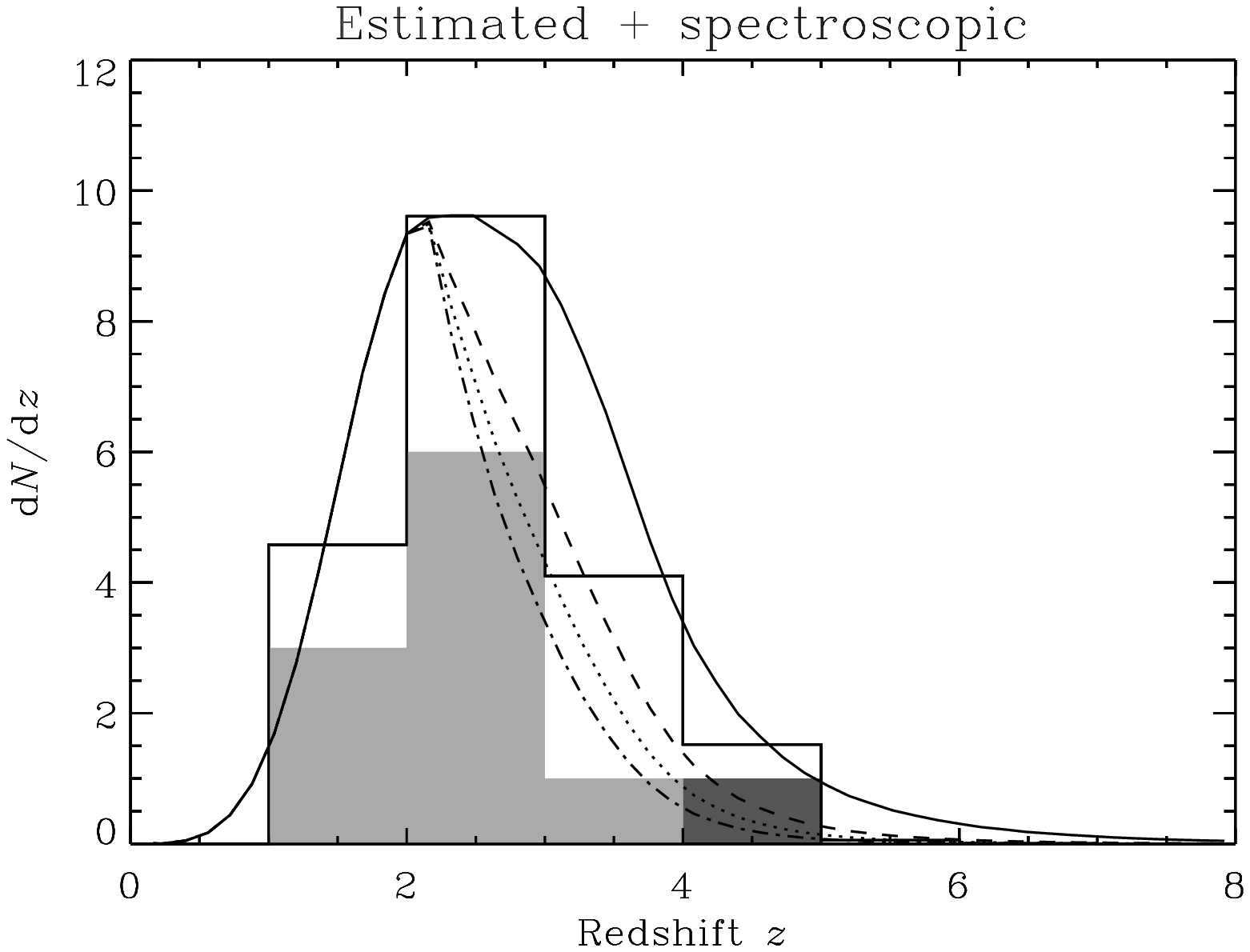}   
    \end{minipage}
  \end{center}
\caption{The binned redshift distribution of sources in the top-decade
(i.e. with $\log_{10} L_{151} \geq 27.61$), constructed by taking the
redshift probability density functions of Section~\ref{sec:pdf} into
account, rather than just the best-fitting redshift estimates (see
Fig~\ref{fig:redshift}), as described in
Section~\ref{sec:pdfdistribution}. The bin width is $\Delta z =
1$. Each source contributes to a number of bins, with an amount
proportional to the probability of the source lying within that bin.
The available spectroscopic redshifts are also taken into
consideration.  These contribute to a single bin each and are
represented by the light shaded region. The solid line shows Jarvis et
al. (2001c) model C predictions ($\eta = -0.06$, constant co-moving
space density) for the redshift distribution
(Section~\ref{sec:predicted}). The dashed, dotted and dot-dashed lines
show the same model predictions but for declining co-moving space
densities with $\eta = -2.0$, $\eta = -3.0$ and $\eta = -4.0$,
respectively. The dark shaded region represents 6C**0737+5618, for
which the redshift estimate is obtained based on extrapolation of the
$K-z$ relation, and a probability density function was not derived
(see also the caption of Table~\ref{tab:estimates_median}).  This
source contributes to a single bin, the $4 <z < 5$ bin.}
  \label{fig:redshift.pdf}
\end{figure}

\subsubsection{Taking the redshift probability distributions into account }
\label{sec:pdfdistribution}

We now construct the redshift distribution of the top-decade sources
by taking into account the redshift probability distribution of each
source (derived from the $K-z$ diagram), rather than just using the
best-fitting redshift estimate. In practice, since the probability
density functions (derived in Section~\ref{sec:pdf}) are normalised to
unity, when the binned redshift distribution is constructed each
source contributes to a number of redshift bins, with an amount
proportional to the probability of the source lying within that
bin. For each source, we calculate also the redshift which corresponds
to a luminosity of $\log_{10}\, L_{151} \geq 27.61$, which is the
lower limit of the top-decade, and only take into account
contributions which lie above that redshift.  All the sources are
considered, i.e. we include also those sources whose best-fitting
redshift estimate would put them below the lower-limit of the
top-decade in luminosity, and consider the contribution from the
high-redshift tail of their probability distributions. The
spectroscopic redshifts, when available, are also taken into
consideration. The sources with spectroscopic redshifts contribute to
a single bin each, if their luminosity lies above the lower limit of
the top-decade (see Table~\ref{tab:top-decade}). 6C**0737+5618, the
source which is not detected in $K$-band and does not have a
spectroscopic redshift, contributes also to a single bin ($4 < z <
5$). The resulting redshift distribution is presented in
Fig.~\ref{fig:redshift.pdf}.

\begin{table}
\begin{center}
\begin{tabular}{l|ll|lc|l}
\hline
\mc{1}{c|}{} & \mc{2}{|c|}{} & \mc{2}{|c|}{Omitting} & \mc{1}{c}{Notes}\\
\mc{1}{c|}{} & \mc{2}{|c|}{} & \mc{2}{|c|}{6C**0737+5618} & \mc{1}{c}{}\\
\mc{1}{c|}{$\eta$} & \mc{1}{|c}{$\chi^{2}$} & \mc{1}{c|}{$p$} & \mc{1}{c}{$\chi^{2}$} & \mc{1}{c|}{$p$} &\\
\hline
-0.06 & 1.2 & 0.54 & 2.4 & 0.36 & constant space density\\
-2.0  & 1.4 & 0.24 & 0.8 & 0.36 & declining space density\\
-2.5  & 2.4 & 0.12 & 1.6 & 0.21 & declining space density\\
-3.0  & 3.7 & 0.05 & 2.6 & 0.11 & declining space density\\
-3.5  & 5.2 & 0.02 & 3.8 & 0.05 & declining space density\\
-4.0  & 6.8 & 0.009& 5.1 & 0.02 & declining space density\\
\hline
\end{tabular}
\end{center}
{\caption{\label{tab:chi-square} The results of the chi-squared
goodness-of-fit test to compare the model and data redshift
distributions of Fig.~\ref{fig:redshift.pdf}. The first column lists
the value of the parameter $\eta$, which is the power-law exponent
describing the high-redshift co-moving space density (see
Eq.~\ref{eqn:rhozC}). Columns two and three give the values of
chi-square $\chi^{2}$, and the probability $p$ of obtaining that value
of $\chi^{2}$ or greater, when including 6C**0737+5618 in the
analysis.  Columns four and five list the same values, when omitting
this source.}}
\end{table}

The distribution derived from the data appears to be in very good
agreement with the model distribution. This is confirmed by the
application of the chi-squared goodness-of-fit test, which gives
\mbox{$\chi^{2} = 1.2$}, with a probability $p = 0.54$ of obtaining
this, or greater than this value of $\chi^{2}$.

We perform also the same analysis by omitting 6C**0737+5618 in the $4
\leq z \leq 5$ bin. This source is unidentified and it is debatable
that its \kband magnitude limit ($K \,\,\gtsim\,\, 21$~mag in a
8-arcsec aperture at the 3$\sigma$ level) implies, by extrapolation of
the $K-z$ diagram, a very high redshift ($z > 4$). It is also possible
that this source, like 6C**0935+4348, is at a lower redshift ($z
\simeq 2-3$) but is perhaps at a very early stage of its formation
and/or highly obscured, or is simply underluminous (see discussion in
Section~\ref{sec:comparison}).  However, a very high-redshift cannot
be ruled out.

Omitting 6C**0737+5618 from our analysis does not change the results
significantly.  We find $\chi^{2} = 2.4$, with $p = 0.36$.  Again, we
conclude that the data are consistent with the constant co-moving
space density model of Jarvis et al. (2001c).

Finally, we compare the data with the predictions from five declining
co-moving space density models, using Jarvis et al. (2001c) model C
with: $\eta = -2.0$ (dashed line in Fig.~\ref{fig:redshift.pdf}),
$\eta = -2.5$, $\eta = -3.0$ (dotted line in
Fig.~\ref{fig:redshift.pdf}), $\eta = -3.5 $ and $\eta = -4.0$
(dot-dashed line in Fig.~\ref{fig:redshift.pdf}). The results of the
chi-squared goodness-of-fit test are listed in
Table~\ref{tab:chi-square}. It can be seen that, although the data are
also consistent with moderate declines by factors of 2 to 3, declines
by a factor of 3.5 and 4.0 can be excluded at the $\sim 2 - 3\sigma$
level.

\subsubsection{The limitations of a filtered sample}

As discussed by Jarvis (2000), filtered samples such as 6C** have
limitations when it comes to confirming a decline in co-moving space
density.  The major problem is that the lack of sources at a given
redshift may not be due to a decline in their space density but to
imperfections in the filtering technique.  For example, with respect to
the 6C* sample, Jarvis (2000) estimated that the angular size
selection is filtering out an increasingly large fraction of the
sources with redshift: from $\sim 20$\% to $\sim 30$\% between $z = 0$
and $z \sim 5$; and $\sim 30-50$\% beyond $z > 5$.  Jarvis (2000)
concluded that samples with similar filtering criteria to that of 6C*,
and in particular with flux-density limits similar to that of 6C**,
are only able to confirm roughly constant or increasing co-moving
space density at high redshifts. The presence of a decline would be
difficult to interpret due to the uncertainties introduced by the
filtering criteria. However, the filtering is helpful in placing
strong lower limits on any decline.

\section{Summary}\label{sec:conclusions}

A method of redshift estimation, based on the $K-z$ diagram of the
3CRR, 6CE, 6C* and 7CRS radio galaxies has been developed.  Redshift
probability density functions are derived for all of the 6C** sources
which are identified with a near-infrared counterpart, i.e. for 66 of
the 68 members of the sample.  Comparison of the resulting redshift
estimates with the subset of spectroscopic redshifts shows that our
method is fairly robust whenever emission-line and/or non-stellar
contributions to the $K$-magnitudes can be neglected. The estimated
redshift distribution has a median redshift of $z_{\rm med} \simeq 1.6
$.  However, we find that the quasars have their redshifts
significantly under-estimated by our method. This is explained by the
fact that the method is based on the $K-z$ relation, which is only
valid for radio galaxies. Removing the quasars from the distribution
results in a median estimated redshift of $z \approx 1.7$. This is
similar to that of the 6C* sample ($z_{\rm med} \approx 1.9$) and is
significantly higher than that of unfiltered, complete surveys at the
same flux density limit.  We conclude that the filtering criteria were
effective in biasing the 6C** sample to objects at high-redshift.

The redshift distribution of the most luminous sources in the 6C**
sample is compared with the predictions of the steep-spectrum RLF
model of Jarvis et al. (2001c).  We find that the 6C** data is
consistent with a constant co-moving space density at $z
\,\,\gtsim\,\, 2.5$, and moderate declines by factors of $\sim$ 4 can
be excluded at the $\sim 2 - 3\sigma$ level. Although Jarvis et
al. (2001c) excluded these declines at the $\sim 4\sigma$ level, the
additional data from 6C** provide an independent measure. Thus, the
two independent studies are in quantitative agreement with the result
that any decline at high redshift is shallow.

We note that our result is based on a redshift distribution which is
uncertain for the following reasons: (i) a significant fraction of the
sample is not identified spectroscopically; (ii) the method of
redshift estimation relies on the limited statistics of the 3CRR, 6CE,
7CRS and 6C* radio galaxies at $z > 3$ (the scatter in the $K-z$
diagram at these high redshifts is still poorly defined); and (iii)
the redshift estimates of quasars are systematically under-estimated
by our method. Most of these are likely to lead to an under-estimate
of the true median redshift of the complete sample. Thus, although
spectroscopically incomplete, with the 6C** sample we have additional
strong constraints on the high-redshift space density, with a sample
that increases the number of powerful steep-spectrum sources, from
complete samples at $z > 2$, by a factor of $\sim$ 2.

The work presented here could be significantly improved by obtaining
spectroscopic redshifts for a larger fraction of sources in the 6C**
sample. This would be particularly important for the faintest sources
($K \,\,\gtsim\,\, 19$\,mag), since these are the most probable $z > 2$
candidates. With spectra of these sources we should be able to obtain
a tighter constrain on the co-moving space density at $z > 2$.

\section*{ACKNOWLEDGEMENTS} 
We thank Isobel Hook and Ross McLure for very useful comments.  MJC
acknowledges the support from the Portuguese Funda\c{c}\~{a}o para a
Ci\^{e}ncia e a Tecnologia, and the receipt of a NOVA Marie Curie
Early Stage Training Fellowship.  She also gratefully acknowledges the
generous hospitality of the Institute for Computational Cosmology and
the Extragalactic Cosmology Research Group, at Durham University.  KMB
acknowledges the Royal Society for a University Research Fellowship.

\end{document}